\documentclass[preprint,12pt]{elsarticle}

\usepackage{amssymb}
\usepackage{amsmath}
\usepackage{subcaption}
\usepackage{caption}
\usepackage{url}

\makeatletter
\def\ps@pprintTitle{%
  \let\@oddhead\@empty
  \let\@evenhead\@empty
  \let\@oddfoot\@empty
  \let\@evenfoot\@empty
}
\makeatother

\begin{document}

\begin{frontmatter}



\title{Enhancing industrial microalgae production through Economic Model Predictive Control}

\author[UAL]{Pablo Otálora} 
\author[NTNU]{Sigurd Skogestad}
\author[UAL]{José Luis Guzmán}
\author[UAL]{Manuel Berenguel}

\affiliation[UAL]{organization={Department of Informatics, University of Almería},  addressline={CIESOL, ceiA3}, city={Almería}, postcode={04120}, country={Spain}}

\affiliation[NTNU]{organization={Department of Chemical Engineering, Norwegian University of Science and Technology},city={Trondheim}, postcode={NO-7491}, country={Norway}}

\begin{abstract}
The industrial production of microalgae is an important and sustainable process, but its actual competitiveness is closely related to its optimization. The biological nature of the process hinders this task, mainly due to the high nonlinearity of the process along with its changing nature, features that make its modeling, control and optimization remarkably challenging. This paper presents an economic optimization framework aiming to enhance the operation of such systems. An Economic Model Predictive Controller is proposed, centralizing the decision making and achieving the theoretical optimal operation. Different scenarios with changing climate conditions are presented, and a comparison with the typical, non-optimized industrial process operation is established. The obtained results achieve economic optimization and dynamic stability of the process, while providing some insight into the priorities during process operation at industrial level, and justifying the use of optimal controllers over traditional operation.
\end{abstract}



\end{frontmatter}


\section{Introduction}\label{sec:introduction}

In the face of growing global concerns about climate change, resource depletion, and environmental degradation, there is an urgent need for sustainable technologies that can mitigate anthropogenic impacts while supporting economic and social development. Conventional industrial practices, and particularly those related to energy production, agriculture, and waste management, are the main contributors to greenhouse gas emissions and ecological imbalance. As governments and industries strive to transition to circular and low-carbon economies, the search for innovative and efficient biotechnological solutions has become a priority.

In this context, microalgae have emerged as a promising solution. These microorganisms have fast growth rates, high photosynthetic efficiency, and capacity to capture and utilize carbon dioxide, offering a sustainable platform for a variety of industrial applications \cite{acien2017microalgae}. Their production does not compete directly with arable land or freshwater resources, making them particularly appealing in the context of food and water security. Moreover, their remarkable versatility has led to applications in sectors such as energy, agriculture, pharmaceuticals, and environmental remediation \cite{acien2017economics}.

In the energy sector, their high lipid content enables the production of biodiesel as a renewable alternative to fossil fuels \cite{lam2012microalgae}. In agriculture, microalgal biomass can be converted into biofertilizers and animal feed, supporting more sustainable farming practices \cite{guo2020microalgae}. Their rich profile of bioactive compounds makes them valuable in the pharmaceutical and nutraceutical industries \cite{rahman2020food}. In addition, their ability to absorb nutrients and sequester $\mathrm{CO_2}$ has made them increasingly useful for wastewater treatment and the mitigation of flue gases \cite{razzak2017biological}. These diverse applications, combined with their scalability and adaptability, position microalgae as a key biotechnological resource to advance sustainable development.

Microalgae production systems can be broadly classified into closed and open photobioreactors. Closed photobioreactors offer precise control over environmental conditions, reduce contamination risks, and can achieve high biomass productivity; however, they are typically expensive to build and operate, and their scalability is limited by complex engineering and maintenance requirements \cite{narala2016comparison}. In contrast, open reactors, and more specifically raceway reactors, are much more cost effective, easier to scale, and suitable for outdoor operation, making them the most widely adopted configuration for large-scale applications. Despite their lower productivity and susceptibility to contamination and environmental fluctuations, raceway reactors strike a practical balance between economic feasibility and operational simplicity \cite{posten2009design}. Their design makes them especially suitable for industrial-scale microalgae production for biofuels, wastewater treatment, and low-cost biomass.

Despite their widespread use, raceway reactors face several limitations that can hinder microalgal productivity and overall system efficiency. One of the most relevant ones is the open nature of raceway systems, which makes it challenging to maintain ideal growth conditions consistently, due to changing outdoor conditions. This, coupled with the nonlinear, highly changing dynamics of the biological system, makes it difficult to keep a high growth rate on the culture \cite{caparroz2024novel,guzman2025microalgae}. In this context, advances in process control and dynamic optimization offer promising pathways to mitigate these issues and improve the competitiveness of the process \cite{guzman2020modelado, guzman2025microalgae, bernard2016modelling}.

There are many contributions that have sought solutions to each of these problems. Some authors have succeeded in developing models that describe different aspects of the dynamics of the process. Certain works, like \cite{bernard2016modelling,solimeno2019bio_algae,sanchezzurano2021abaco,nordio2024abaco2,casagli2021alba} have focused on modeling the biological part of the process, describing how microaglae biomass grows depending on culture conditions. Other works have developed engineering models that describe the evolution of the variables of the system, like in \cite{fernandez2014first, rodriguezmiranda2025acomprehensive}, where a first principles model describing pH and dissolved oxygen is presented. In \cite{rodriguezmiranda2021anew} a thermal model of a raceway reactor is proposed. Other works, like \cite{caparroz2024novel,otalora2021dynamic} focus on data-driven approaches instead on first principles model, such as regression trees and artificial neural networks, for pH prediction.

Further works have focused on the control of the most influential variables in the productivity of the process, typically pH and dissolved oxygen (DO), but also temperature. In \cite{rodriguezmiranda2019daytime,rodriguezmiranda2020diurnal,pawlowski2016event}, PID-based control strategies for pH regulation in raceway reactors are proposed, combined with event-based control approaches. In \cite{mairet2015adaptive}, authors proposed a strategy for light attenuation control, influencing the productivity of the system. \cite{rodriguezmiranda2021indirect} presented a temperature control strategy by manipulating the water level of the reactor. In \cite{pataro2023alearning}, a learning-based model predictive controller is proposed for pH control in freshwater and wastewater raceway reactors, and particularly those related with economic optimization solutions.

Few authors have focused on achieving economic or productive dynamic optimization of the process. In \cite{delahozsiegler2012optimization}, authors propose a nonlinear model-based strategy for optimizing microalgae productivity in closed reactors. In \cite{pfaffinger2016model}, another model-based strategy is used for optimizing productivity in flat-pannel reactors. In \cite{fernandez2016hierarchical}, MPC and PID layers are combined in a hierarchy to optimize productivity in closed reactors by manipulating the pH setpoint. \cite{dewasme2017microalgae} proposed a model-free extremum-seeking control approach to optimize productivity in closed reactors. \cite{bernard2022optimal} performed a theoretical analysis of the productivity optimization, combined with a nonlinear control, to optimize growth a microalgae reactor. \cite{ifrim2022model} proposed an optimal controlled for productivity optimization in closed photobioreactors. In view of this, it is noticeable that there is a gap in the literature regarding the optimization of open reactors, and particularly in outdoors conditions and at industrial or semi-industrial scale.

In this paper, the objective is to develop and propose a dynamic optimization framework for the operation of a raceway reactor, with a focus on economic performance. By integrating the dynamics of the process with economic criteria, the aim is to enhance the overall efficiency and profitability of microalgae production in open systems. This approach will explicitly consider operational costs, resource utilization, and potential revenue from biomass production. Through this framework, the paper also seeks to provide valuable insight into key operational priorities and trade-offs that should guide decision making in large-scale microalgal processes. Ultimately, the proposed strategy serves as a step toward more economically viable and sustainable industrial applications of microalgae.

This paper is organized as follows. Section \ref{sec:system} provides a detailed description of the raceway reactor system studied, including the underlying first-principles model used to represent the process dynamics. Section \ref{sec:EMPC} introduces the economic model predictive control (EMPC) framework. 

Section \ref{sec:resultados} presents the simulation results, starting with a baseline non-optimized scenario. It then introduces the definition of the economic cost function used to assess performance and the operational constraints applied during the optimization process, describes the EMPC tuning procedure, and highlights the performance improvements achieved under the optimized strategy. Additionally, it explores various approaches for forecasting and incorporating future disturbances. Section \ref{sec:conclusiones} concludes the paper by summarizing the key findings and outlining potential directions for future research.

\section{System description}\label{sec:system}

This section provides an overview of the system considered for this study, including a detailed description of the microalgae raceway reactor and its instrumentation, the first-principles model used to represent the process dynamics, and the economic cost function and operational constraints that form the basis for the optimization framework.

\subsection{Microalgae raceway reactor}\label{subsec:reactor}
\begin{figure}[ht]
    \centering
    \includegraphics[width=0.8\textwidth]{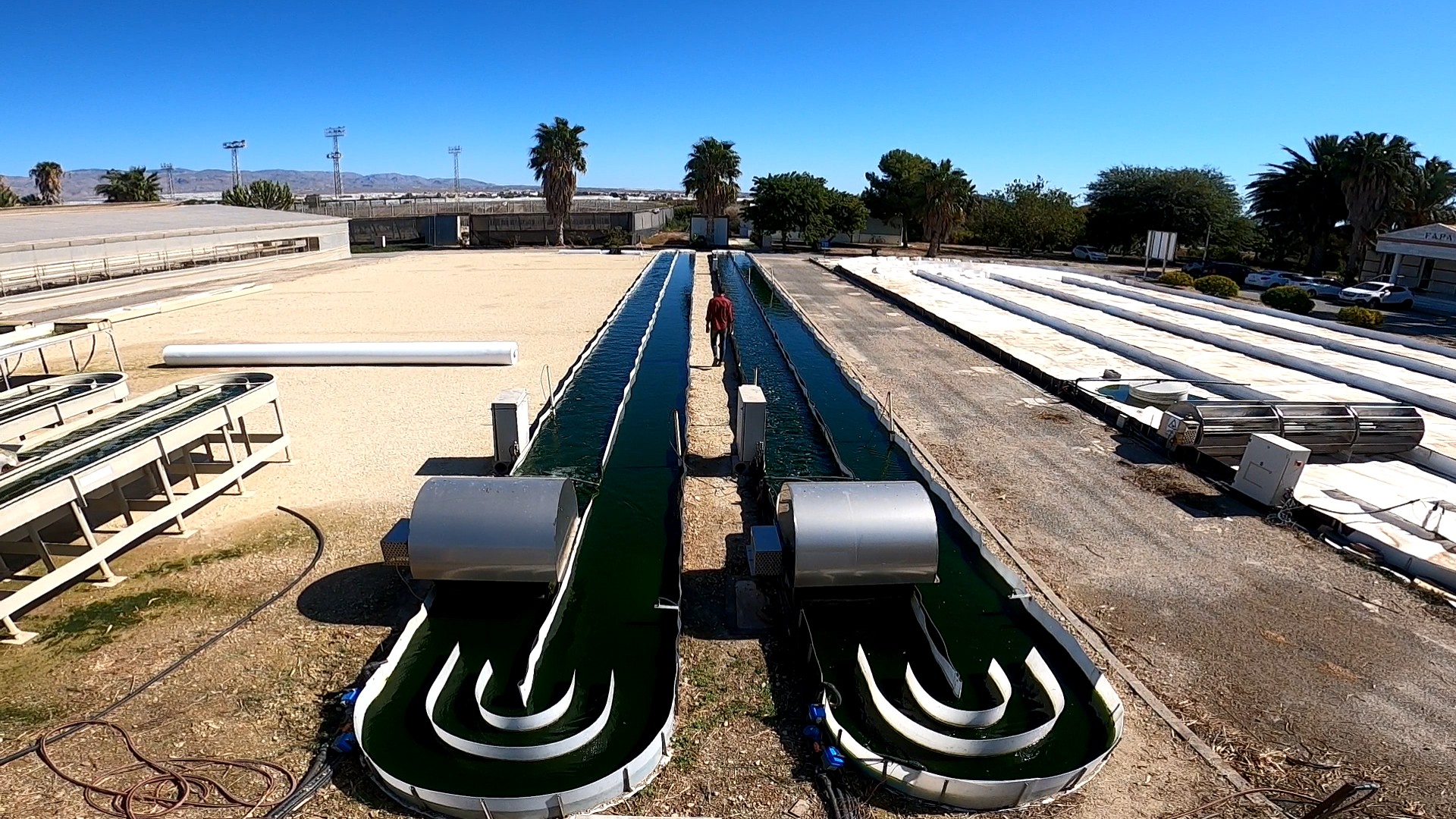}
    \caption{Semi-industrial scale raceway photobioreactor located in the CIESOL research center at the IFAPA facilities.}
    \label{fig:reactor}
\end{figure}

The system studied in this work is a raceway photobioreactor located at the IFAPA research center, near the University of Almería (see Figure \ref{fig:reactor}). The reactor is an open system consisting of two parallel channels, each 40 meters long, 1 meter wide, and 30 centimeters deep, connected at their ends by 180º bends to form a continuous loop. The circulation of the culture medium is driven by a paddle wheel located near one of the bends, which maintains a constant linear flow velocity and promotes turbulence to enhance mixing within the reactor.

The reactor is also equipped with a sump located immediately downstream of the paddle wheel, where carbon dioxide injection and air bubbling are carried out. These inputs, as will be discussed later, are used to regulate the culture conditions. The outflow of culture medium is managed by a pump, while the inflow, also called dilution, is controlled via an electrovalve that introduces fresh, nutrient-rich, medium into the system. The harvested medium is then directed to a post-processing stage, where it is prepared for use in other applications.

The system is instrumented with pH, dissolved oxygen, and temperature sensors positioned at two key locations: immediately downstream of the sump, in the direction of the medium flow, and just before the second bend of the reactor loop. The latter location, being the point most distant from the air and $\mathrm{CO_2}$ injection site, typically presents the greatest challenge for maintaining stable culture conditions and is therefore often selected as the primary control target. Consequently, sensor data from this point are used for model development. Additional instrumentation includes sensors for water level and biomass concentration, as well as external environmental variables such as incident solar radiation, ambient temperature, relative humidity, and soil temperature.

\begin{figure}[ht]
    \centering
    \includegraphics[width=0.8\textwidth]{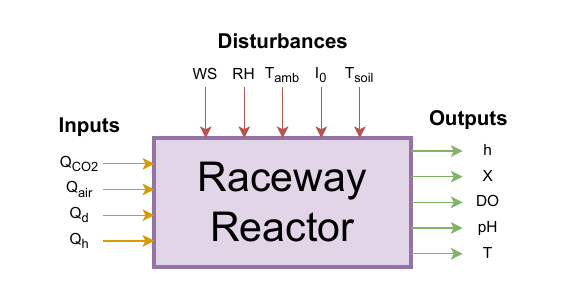}
    \caption{Inputs (MVs), disturbances and outputs (CVs) of the system.}
    \label{fig:flowchart}
\end{figure}

Figure \ref{fig:flowchart} illustrates the inputs (manipulated variables), disturbances, and outputs (controlled variables) of the system. The manipulated inputs consist of various flow rates: air bubbling ($Q_{air}$), $\mathrm{CO_2}$ injection ($Q_{CO_2}$), dilution or inflow rate ($Q_d$), and harvesting or outflow rate ($Q_h$). The primary external disturbances include wind speed ($WS$), relative humidity ($RH$), ambient temperature ($T_{amb}$), global irradiance ($I_0$), and soil temperature ($T_{soil}$). The main system outputs, all of which are measurable, comprise the water level ($h$), biomass concentration ($X$), dissolved oxygen ($DO$), pH, and water temperature ($T$). The system features two additional states, those being $\mathrm{CO_2}$ concentration ($\mathrm{[CO_2]}$) and Total Inorganic Carbon (TIC) concentration ($\mathrm{[TIC]}$), which cannot be measured in real time.

Sensor data is recorded continuously every second, 24 hours a day throughout the year. These data serve a dual purpose: they are used both for model calibration and as input to the model during simulations, enabling the recreation of realistic environmental conditions for testing the developed control algorithms.

The modeled reactor operates using clean water supplemented with externally supplied nutrients as the culture medium. The microalgal species cultivated is \textit{Scenedesmus almeriensis}, known for its high tolerance to a broad range of environmental conditions and its suitability for open reactor systems \cite{sanchez2021wastewater}. This species thrives particularly well within a pH range of 7 to 9.5 and temperatures between 10 and 40$^o$C.

\subsection{First-principles model}\label{subsec:model}
The system model used for process optimization and simulation is based on a combination of previously published models \cite{sanchezzurano2021abaco}, \cite{rodriguezmiranda2021anew}, \cite{fernandez2016dynamic} and \cite{rodriguezmiranda2025acomprehensive}, with specific modifications introduced to suit the objectives of this study. The model is structured into three main components: the biological model, which captures the dynamics of biomass growth; the dynamic model which represents the behavior of dissolved oxygen and pH; and the thermal model, which describes the reactor heat exchange mechanisms and the resulting evolution of the water temperature. A well-mixed reactor is assumed throughout the modeling. This model is used both for the design of the proposed optimization strategy and as a simulation platform to evaluate different operational scenarios.

\subsubsection{Biological model}\label{subsubsec:biologico}
The biological model describes the temporal evolution of biomass concentration within the reactor. It is based on a mass balance that accounts for two main contributions: the net biomass gain from photosynthetic growth and the losses due to dilution and evaporation. This relationship comes from the biomass mass balance, presented in Equation \eqref{eq:dB}, where $X(t)$ represents the biomass concentration, $\mu(t)$ is the specific growth rate, $Q_h$ is the harvesting flow rate, and $V(t)$ is the total volume of water in the reactor. 

\begin{equation}\label{eq:dB}
    \frac{d(X(t)\cdot V(t))}{dt}=X(t)\cdot V(t)\cdot\mu(t)-Q_h(t)\cdot X(t)
\end{equation}

This expression can be rearranged considering the water mass balance, which is presented in Equation \eqref{eq:dh}, where $A$ is the area of the reactor, $h(t)$ is the water level (being $A\cdot h(t)=V(t)$), $Q_d(t)$ denotes the dilution flow rate, and $g_s(t)$ is the evaporation rate. Specifically, water exits the reactor as a result of harvesting and evaporation, while it enters through the dilution flow. Among these terms, the evaporation rate is the most complex to model.

\begin{equation}\label{eq:dh}
    A\frac{dh(t)}{dt}=Q_d(t)-Q_h(t)-g_s(t)
\end{equation}

The evaporation rate depends on several environmental and operational variables: water temperature, ambient temperature, relative humidity, wind speed, and air flow rate. It is calculated using Equation \eqref{eq:gs} \cite{tang2004comparative}, which incorporates the gas-liquid transfer coefficient $K_l$, which depends on wind speed $WS$ and air flow rate $Q_{air}$, the saturation vapor pressure at water temperature $e_w$, the ambient vapor pressure $e_a$, and the latent heat of vaporization $L_v$. These interactions make evaporation a dynamic and sensitive component of the water balance in open reactors.

\begin{equation}\label{eq:gs}
    g_s(t)=K_l(WS,Q_{air})\cdot A\cdot(e_w(T)-e_a(T_{amb},RH))\cdot\frac{1}{L_v(T)}
\end{equation}

From these expressions, Equation \eqref{eq:dX} can be reached, describing the main dynamics of biomass concentration.

\begin{equation}\label{eq:dX}
    \frac{dX(t)}{dt}=X(t)\cdot\left(\mu(t)-\frac{Q_d(t)-g_s(t)}{V(t)}\right)
\end{equation}

An increase in biomass concentration can result from either water evaporation, which concentrates the culture, or from actual microalgal growth. The growth rate depends on the environmental and nutrient conditions within the reactor and is described by Equation \eqref{eq:mu}, as proposed in \cite{sanchezzurano2021abaco}. In this expression, the term $\mu_{Iav}$ represents the maximum potential growth rate, which is a function of the average radiation available to the culture ($I_{av}$). This potential is modulated by a series of normalized weighting factors $\mu_i$, each ranging from 0 to 1, which account for the influence of key variables: temperature ($T$), pH, dissolved oxygen ($DO_2$), and the concentrations of carbon dioxide ($\mathrm{CO_2}$), nitrates (N), and phosphates (P). Furthermore, the model includes a maintenance term $m(t)$, which reflects biomass losses due to endogenous respiration and decay. This term is negative with respect to growth and also depends on the radiation received by the culture.

\begin{equation}\label{eq:mu}
    \mu(t)=\mu_{Iav}(t)\cdot\mu_T(t)\cdot\mu_{pH}(t)\cdot\mu_{DO_2}(t)\cdot\mu_{CO_2}(t)\cdot\mu_N(t)\cdot\mu_P(t)-m(t)
\end{equation}

In a freshwater reactor such as the one studied in this work, nitrates and phosphates are supplied through the dilution flow and can be assumed to be in excess, allowing their corresponding growth terms to be approximated as 1. A similar assumption can be made for $\mathrm{CO_2}$, which is actively injected during operation. As a result, analysis can focus on the remaining growth limiting factors to better understand the behavior of a control system aimed at optimizing the process.

The influence of the average irradiance on the growth rate is described by Equation \eqref{eq:muiav}, where $\mu_{max}$ is the maximum specific growth rate, $I_k$ is the constant of irradiance, representing the level of irradiance at which half of $\mu_{max}$ is achieved and $n$ is a form parameter. This formulation captures the typical saturation behavior observed in photosynthesis, where increasing irradiance initially enhances growth, but eventually leads to diminishing returns as the growth rate approaches a maximum.

\begin{equation}\label{eq:muiav}
    \mu_{Iav}(t)=\mu_{max}\cdot\frac{I_{av}^n(t)}{I_{av}^n(t)+I_k^n}
\end{equation}

Although excessively high irradiance can lead to photoinhibition, thus reducing microalgal growth, this phenomenon is not considered in the current model. The average irradiance available to the microalgae is determined by the global irradiance at the reactor surface ($I_0$), the biomass concentration and the water level, as described by Equation \eqref{eq:Iav}. In this expression, $K_a$ denotes the biomass extinction coefficient. Although $I_{av}$ is directly proportional to $I_0$, it decreases as biomass concentration and water depth increase, due to increased light attenuation within the culture medium.

\begin{equation}\label{eq:Iav}
    I_{av}(t)=\frac{I_0(t)}{K_a\cdot X(t)\cdot h(t)}\cdot\left(1-e^{-K_a\cdot X(t)\cdot h(t)}\right)
\end{equation}

\begin{figure}[ht]
    \centering
    \includegraphics[width=0.8\textwidth]{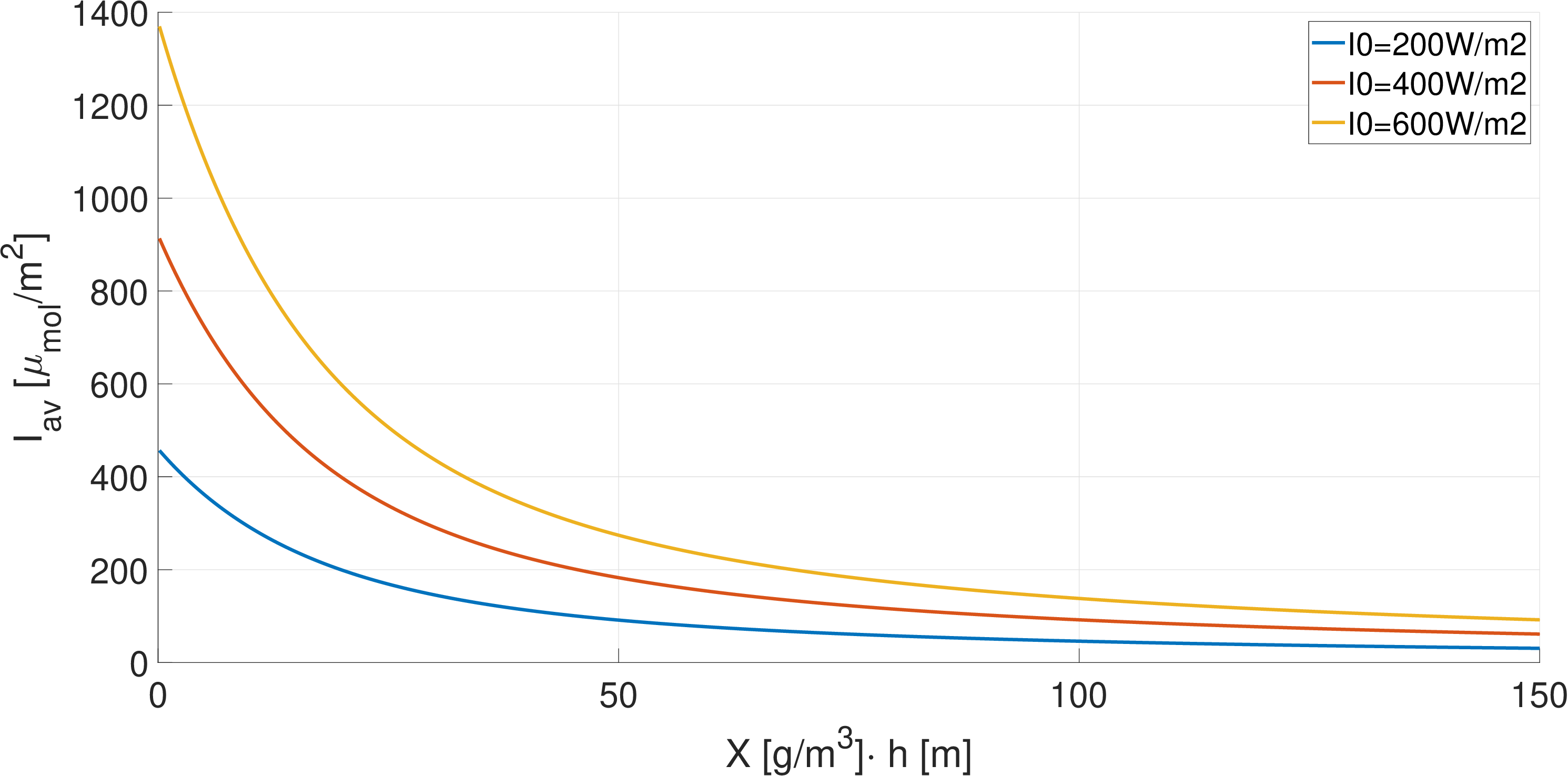}
    \caption{Average irradiance received by microalgae depending on the biomass concentration from Equation \eqref{eq:Iav}.}
    \label{fig:Iav_Cb}
\end{figure}

This attenuation effect significantly affects the growth of microalgae, as is evident from the formulation of the growth rate in Equation \eqref{eq:muiav}. Since biomass concentration and water level appear multiplicatively in the expression for $I_{av}$, increases in either variable lead to similar reductions in light availability and thus in growth. This relationship is illustrated in Figure \ref{fig:Iav_Cb}, which shows how $I_{av}$ decreases with increasing biomass concentration at constant water depth. A similar trend would be observed if the biomass concentration was kept constant while varying the water level.

\begin{figure}[ht]
    \centering
    \includegraphics[width=0.8\textwidth]{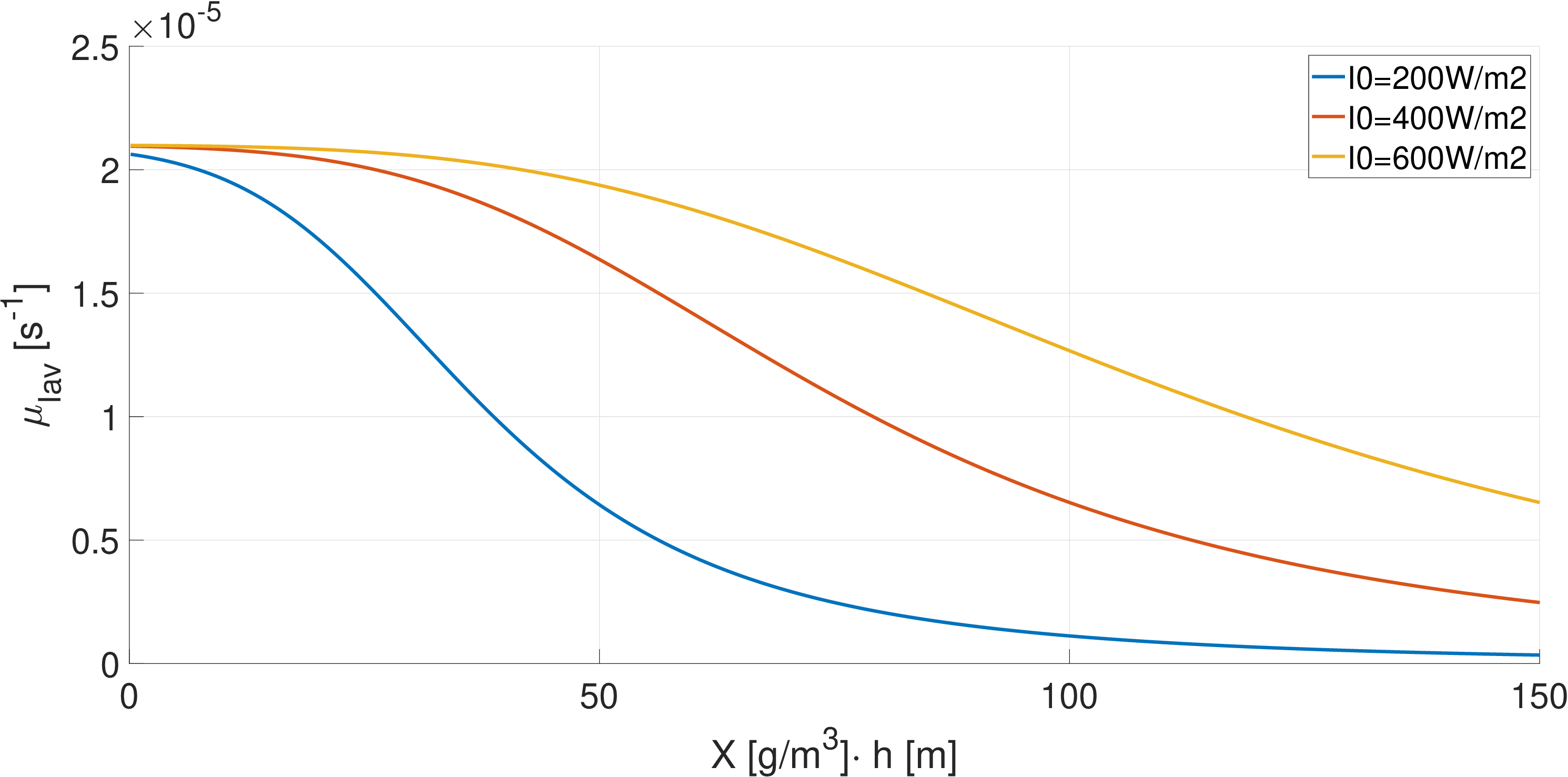}
    \caption{Growth rate depending on the radiation and biomass concentration from Equations \eqref{eq:muiav} and \eqref{eq:Iav}.}
    \label{fig:uIav_Cb}
\end{figure}

Since average irradiance directly influences the growth rate, changes in biomass concentration or water level also impact microalgal growth. Figure \ref{fig:uIav_Cb} illustrates how the growth rate varies with the concentration of biomass under different levels of incident radiation, while keeping the water level constant. As expected, an increase in the concentration of biomass leads to a reduction in the growth rate as a result of the corresponding decrease in light availability within the culture.

The temperature and pH terms in the growth model share a similar mathematical structure: both exhibit a maximum contribution to productivity at an optimal value and drop to zero outside defined minimum and maximum thresholds. These relationships are described by Equations \eqref{eq:muT} and \eqref{eq:mupH}, for temperature and pH, respectively. In the case of temperature, $T(t)$ denotes the actual water temperature, $T_{opt}$ is the optimal temperature for the produced microalgae strain, and $T_{min}$ and $T_{max}$ define the temperature range beyond which growth is completely inhibited. The same nomenclature applies to pH. From these expressions, it follows that maintaining the temperature and pH close to their optimal values (30°C and 8, respectively) is essential to maximize the productivity of microalgae.

\begin{equation}\label{eq:muT}
\resizebox{0.9\hsize}{!}{$
    \mu_T(t)=\frac{(T(t)-T_{max})\cdot(T(t)-T_{min})^2}{(T_{opt}-T_{min})\cdot((T_{opt}-T_{min})\cdot(T(t)-T_{opt})-(T_{opt}-T_{max})\cdot(T_{opt}+T_{min}+2T(t)))}
    $}
\end{equation}

\begin{equation}\label{eq:mupH}
\resizebox{0.9\hsize}{!}{$
    \mu_{pH}(t)=\frac{(pH(t)-pH_{max})\cdot(pH(t)-pH_{min})^2}{(pH_{opt}-pH_{min})\cdot((pH_{opt}-pH_{min})\cdot(pH(t)-pH_{opt})-(pH_{opt}-pH_{max})\cdot(pH_{opt}+pH_{min}+2pH(t)))}
    $}
\end{equation}

The term associated with dissolved oxygen differs from those associated with temperature and pH, as it follows a distinct functional form, shown in Equation \eqref{eq:muDO}. In this case, the growth contribution is close to 1 at low concentrations of dissolved oxygen and decreases progressively, reaching zero once the critical threshold, $DO_{2,max}$, is exceeded. This threshold varies according to the specific microalgal strain. $m_{DO_2}$ serves as a form parameter, determining the degree of decline in growth as DO approaches inhibitory levels.

\begin{equation}\label{eq:muDO}
    \mu_{DO_2}(t)=1-\left(\frac{DO_2(t)}{DO_{2,max}}\right)^{m_{DO_2}}
\end{equation}

From this expression, it can be inferred that maintaining dissolved oxygen at low concentrations is essential to support optimal microalgal growth, as excessive DO levels can inhibit productivity. Therefore, controlling DO below the strain-specific threshold is a key aspect of process optimization.

The respiration term $m(t)$, shown in Equation \eqref{eq:m}, follows a form similar to that of $\mu_{Iav}(t)$, but includes an additional constant term to ensure that its value never reaches zero. Like the growth term, it also depends on the radiation received by the culture, but its response is more moderate. As a result, under illuminated conditions, the growth contribution is generally expected to exceed the respiration losses, ensuring a net increase in biomass.

\begin{equation}\label{eq:m}
    m(t)=m_{min}+m_{max}\cdot\frac{I_{av}^{n_m}(t)}{I_{av}^{n_m}(t)+I_{km}^{n_m}}
\end{equation}

\subsubsection{Dynamic model for pH and dissolved oxygen}\label{subsubsec:engineering}

The second part of the model was originally developed in \cite{fernandez2016dynamic} and it is available in \cite{rodriguezmiranda2025acomprehensive}. In the present work, this model has been adapted with several modifications to more accurately represent the mechanisms governing mass-transfer processes. To see all the original equations in detail, readers are referred to \cite{fernandez2016dynamic}. The dynamics of dissolved oxygen and pH are described by two fundamental mass balances: the oxygen balance and the carbonate balance, which will be analyzed independently.

The oxygen balance accounts for three primary phenomena: liquid-gas transfer between the reactor and the atmosphere, oxygen production through microalgal photosynthesis, and oxygen input through dilution. The liquid-gas transfer includes a passive component, which is influenced by wind speed, and a forced component resulting from air bubbling in the sump. This bubbling improves gas exchange, thus reducing dissolved oxygen levels, which often exceed 100\% due to intense photosynthetic activity.

The generation of oxygen through photosynthesis is directly proportional to both biomass concentration and growth rate. As a result, dissolved oxygen levels typically rise during midday, when solar irradiance and, consequently, the growth rate are at its peak. In addition, more concentrated cultures lead to greater oxygen production. Finally, dilution introduces fresh medium with an oxygen concentration generally lower than that of the reactor, which tends to reduce the overall DO concentration, albeit to a lesser extent than the effect of air bubbling.

The carbon balance is slightly more complex than the oxygen balance. The first mechanism influencing this balance is the liquid-gas exchange of $\mathrm{CO_2}$ with the atmosphere, which is affected by both wind speed and air bubbling. Given the low atmospheric concentration of $\mathrm{CO_2}$, this mechanism typically results in a net loss of $\mathrm{CO_2}$ from the reactor.

In parallel with oxygen generation, photosynthesis consumes carbon dioxide in proportion to both the biomass concentration and the growth rate. The dilution flow rate also affects the total inorganic carbon content, depending on the carbon composition of the incoming medium.

A fourth key mechanism in the carbon balance is the injection of pure $\mathrm{CO_2}$ into the sump, representing another liquid-gas transfer process. Unlike atmospheric exchange, this injection increases the $\mathrm{CO_2}$ concentration in the reactor and can be used as a control input to regulate pH.

These mechanisms collectively determine the TIC concentration in the reactor, which is closely related to the concentrations of carbon dioxide and hydrogen ions. In general, a higher TIC concentration corresponds to a higher $[H^+]$ concentration, leading to a lower pH, as described by Equation \eqref{eq:pH}. As a result, pH naturally tends to rise during midday due to photosynthetic $\mathrm{CO_2}$ uptake, while air bubbling further increases pH by stripping $\mathrm{CO_2}$. Conversely, $\mathrm{CO_2}$ injection reduces pH, enabling its active control.

\begin{equation}\label{eq:pH}
    pH=-log_{10}([H^+])
\end{equation}

\subsubsection{Thermal model}\label{subsubsec:termico}

The temperature model used in this study is based on the formulation developed in \cite{rodriguezmiranda2021anew}, and thus the main mass balances are described in that original work. Given the relatively minor influence of the manipulable variables on temperature dynamics, this model is not described in as much detail as the biological and dynamic models. However, temperature remains a key factor in microalgal productivity and must be included in the overall analysis of the system. The model describes the dynamics of the water temperature using Equation \eqref{eq:temp}, where $Q_{ac}(t)$ denotes the net heat exchanged with the environment, $C_p$ is the specific heat capacity of the water, $\rho$ is the density of the water and $T_d$ is the temperature of the dilution water, assumed constant. The equation comprises two main terms: one accounts for environmental heat exchange, and the other for heat introduced through dilution. 

\begin{equation}\label{eq:temp}
    \frac{dT(t)}{dt}=\frac{Q_{ac}(t)}{V(t)\cdot C_p\cdot\rho}-\frac{Q_d(t)}{V(t)}\cdot\left(T_d-T(t)\right)
\end{equation}

The term $Q_{ac}(t)$ aggregates several mechanisms of environmental heat transfer, including solar irradiance, radiation, evaporation, convection, and conduction. These processes are strongly influenced by climatic variables such as solar radiation, ambient temperature, relative humidity, and wind speed. Because these variables are continuously monitored and are often predictable through forecasting models, they are treated as predictable disturbances in the optimization framework. For more information about this term, the viewer is invited to read the original paper \cite{rodriguezmiranda2021anew}.

\section{Economic Model Predictive Control}\label{sec:EMPC}

EMPC is an advanced control strategy aimed at optimizing the economic performance of a process, and is formally a particular case of Model Predictive Control (MPC). To properly define EMPC, it is necessary first to understand the foundational principles of its predecessor.

MPC seeks to achieve optimal control by computing a sequence of control actions that drive the system to follow a desired behavior while minimizing a predefined cost function \cite{camacho2007constrained} (see Equation \eqref{eq:MPC}). At any given time step $k$, the optimizer computes the control sequence $u$ that minimizes the integral of the cost function $J$ over a fixed number of future steps, known as the prediction horizon $N_p$. To do this, the optimizer relies on a dynamic model $f$ that captures how the state of the system $x$ evolves over time, based on the current state, planned control actions, and anticipated disturbances $d$. This model is included as an equality constraint in the optimization problem. Additional constraints, typically bounding the manipulable variables or enforcing acceptable ranges for state variables, can also be integrated into the formulation, such as $u_{min}$ and $u_{max}$ on the control signal, or $x_{min}$ and $x_{max}$ on the system states.

\begin{equation}
\begin{aligned}
    \min_{u} \quad & \sum_{i=0}^{N_p} J(k+i \mid k) \\
    \text{S.T.} \quad 
    & \hat{x}(k+i \mid k) = f\left( \hat{x}(k+i-1 \mid k), u(k+i-1 \mid k), d(k+i-1 \mid k) \right), \\
    & u_{\text{min}} \leq u(k+i \mid k) \leq u_{\text{max}}, \quad i = 0, \dots, N_c \\
    & x_{\text{min}} \leq \hat{x}(k+i \mid k) \leq x_{\text{max}}, \quad i = 0, \dots, N_p
\end{aligned}
\label{eq:MPC}
\end{equation}

This optimization problem is formulated and solved in a receding horizon manner. After computing the optimal control sequence, only the first control action is applied to the system. The system is then allowed to evolve over a sampling interval, and the optimization problem is re-evaluated from the new system state. This feedback mechanism ensures that the control strategy is not open-loop, allowing it to compensate for modeling inaccuracies or unanticipated disturbances.

To reduce computational complexity, the optimizer is typically allowed to modify the control inputs only $N_c$ times within the prediction horizon, where $N_c < N_p$. This is known as the control horizon.

MPC offers several advantages over traditional closed-loop control strategies. Ideally, it provides optimal system behavior according to the specified cost function. It can handle nonlinearities effectively by incorporating a dynamic model of the process, and it excels at managing multivariable interactions, as all coupling effects and process constraints are explicitly considered at each step. This is particularly important for state constraints, which are often difficult to manage with conventional control approaches. However, MPC is computationally intensive, especially when non-linear models are used, and heavily model-dependent. Its performance can deteriorate significantly in the presence of modeling errors or unmeasured disturbances.

In standard MPC, the cost function $J$ is typically dynamic, often formulated as a quadratic tracking error relative to a reference trajectory $r$, sometimes with an additional penalty for the control effort:

\begin{equation}
    J(k)=\left(r(k)-\hat{y}(k)\right)^2+\left(u(k)-u(k-1)\right)^2
\label{eq:JMPC}
\end{equation}

In the specific case of an EMPC, the cost function is formulated in economic terms \cite{rawlings2012fundamentals, ellis2014economic}. Account for all factors that influence the economic performance of the process, such as energy consumption, material supplies, and revenues generated from the operation. From the perspective of the automation pyramid, while traditional MPC typically operates at the supervisory or control levels, EMPC extends into higher layers by integrating the optimization layer, an area that, in other approaches, is handled independently and interacts hierarchically with lower-level control systems.

This integration can be particularly challenging, as the optimizer must solve more complex problems than those typically encountered in standard MPC, and do so within relatively short sampling intervals, since it directly controls low-level actuators. At the same time, it must consider relatively long control horizons, as economic dynamics tend to evolve more slowly. This dual focus allows EMPC to capture both fast process dynamics and long-term economic objectives, allowing decisions that are closer to the theoretical optimum while effectively managing the dynamic constraints of the system.

\begin{figure}[ht]
    \centering
    \includegraphics[width=0.5\textwidth]{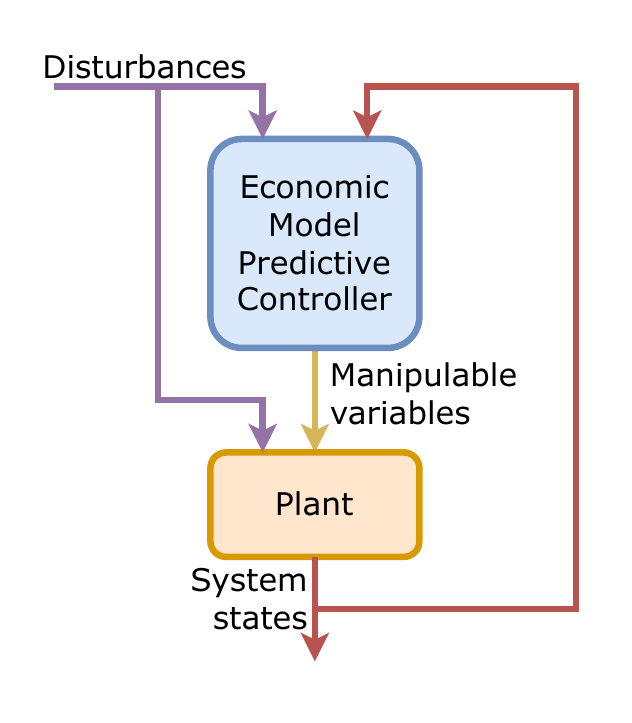}
    \caption{Economic Model Predictive Control diagram.}
    \label{fig:esquema_EMPC}
\end{figure}

The control scheme corresponding to this strategy is illustrated in Figure \ref{fig:esquema_EMPC}. The optimizer receives the current state of the system along with the present and forecasted values of external disturbances. Based on this information, it solves the optimization problem to determine the sequence of control actions that minimize the economic cost function.

\section{Results}\label{sec:resultados}

This section presents the strategies adopted for the economic optimization of the system, along with the results obtained, and a comparison with the traditional mode of operation of the system.

\subsection{Simulation framework}\label{subsec:framework}

Before presenting the results, it is useful to describe the framework in which the different strategies will be simulated. For all cases, real data obtained from the facilities described in Section \ref{subsec:reactor} at different times of the year will be employed. These data serve two main purposes: first, to establish the initial state of the system, using the first recorded value of each state variable; and second, to provide realistic disturbance profiles for the simulated days. The disturbances considered include ambient and soil temperature, global irradiance, wind speed, and relative humidity. These variables were originally sampled every second and subsequently resampled to one-minute intervals using the median.

The model is then integrated from its differential equations using \textit{CasADi}, an open-source tool for nonlinear optimization and algorithmic differentiation \cite{andersson2019casadi}, in \textit{MATLAB}. \textit{CasADi} provides a convenient and efficient framework for time integration of the system dynamics, problem formulation, solution, and cost computation.

The key model parameters used for simulating the system are presented in Table \ref{tab:parametrossistema}. These parameters were calibrated using real system data; however, it is important to note that their values may vary for different microalgae strains or for reactors located in other environments.

\begin{table}[ht]
\centering
\begin{tabular}{r c | r c}
    \hline
    Parameter & Value & Parameter & Value \\
    \hline
    $A$ [m$^2$] & 80 & $K_a$ [m$^2$/g] & 0.1 \\
    $\mu_{max}$ [day$^{-1}$] & 1.8144 & $I_k$ [$\mu$E$\cdot$ m$^{-2}\cdot$ s$^{-1}$] & 120 \\
    $m_{min}$ [day$^{-1}$] & 0.0173 & $n$ [-] & 3 \\
    $m_{max}$ [day$^{-1}$] & 0.0173& $n_m$ [-] & 4 \\
    $T_{min}$ [$^o$C] & 12 & $pH_{min}$ [-] & 4 \\
    $T_{max}$ [$^o$C] & 46 & $pH_{max}$ [-] & 12 \\
    $T_{opt}$ [$^o$C] & 30 & $pH_{opt}$ [-] & 8 \\
    $T_d$ [$^o$C] & 10 & $DO_{2,max}$ [\%] & 500 \\
    $C_p$ [J$\cdot$kg$^{-1}\cdot$K$^{-1}$] & 4184 & $m_{DO_2}$ [-] & 2 \\
    $\rho$ [kg/m$^3$] & 1000 & & \\
    \hline
\end{tabular}
\caption{First principles model parameters.}\label{tab:parametrossistema}
\end{table}

\subsection{Standard operation (benchmark case)}\label{subsec:tradicional}

To properly evaluate the solutions proposed in this work, it is essential to establish a benchmark for comparison. In this case, the benchmark is the conventional operating strategy of the system, which is primarily based on open-loop control and simple rule-based mechanisms. Demonstrating that the optimizer significantly outperforms this traditional approach provides strong justification for the development of such advanced control strategies and supports the broader adoption of microalgae production technologies.

In traditional setup, each manipulable variable is assigned to a specific task in a decoupled manner. The harvesting flow rate is responsible for extracting a fixed volume of culture each day. The fraction harvested typically varies with the time of year, as biomass growth is strongly dependent on seasonal conditions. In real facilities, this harvesting step is usually carried out manually in the morning, by activating the harvest pump at full capacity for a fixed duration. The proportion of the total volume of the reactor that is harvested and replenished each day is referred to as the dilution rate, typically expressed in units of [day$^{-1}$].

This operation is coupled with the dilution flow, one of the few variables controlled in closed-loop. The dilution is managed through a simple on/off controller that ensures that the culture level does not fall below a predefined threshold. This compensates not only for harvesting losses but also for water loss due to evaporation. However, dilution also reduces the biomass concentration. Therefore, larger harvest volumes lead to increased dilution and further decrease in biomass concentration.

One of the main reasons traditional operations are suboptimal is the open-loop nature of the harvesting strategy. On sunny days, when biomass growth is particularly strong, the system may not be able to harvest as much biomass as it could sustainably. In contrast, on cloudy days, maintaining a constant harvest volume can result in end-of-day biomass concentrations lower than those at the start. If this pattern continues over several days, the performance of the system deteriorates in the medium term, often requiring manual intervention, recovery periods, or even culture reinoculation.

The air flow rate, used to maintain dissolved oxygen at levels conducive to growth, is kept constant at a relatively high value. Although this ensures that oxygen levels remain non-inhibitory, it is energetically inefficient because of the high operating cost of the blower. Finally, the last manipulable variable, the $\mathrm{CO_2}$ flow rate, is used to regulate the pH around 8, which is the optimal value for the cultivated species. There are several strategies to achieve this, but the most common approach in microalgae industry is on/off control.

As this description shows, traditional operation addresses the key challenges of the process using a remarkably simple scheme, requiring minimal feedback beyond pH and water level measurements, along with the signals needed for actuator-level cascade control, which is beyond the scope of this study. Table \ref{tab:tradicional} includes a brief summary of some of the parameters of this mode of operation, namely the dilution rate $D$ (which depends on the time of the year) and the different flow rates.

\begin{table}[ht]
\centering
\begin{tabular}{r c}
    \hline
    Parameter & Value \\
    \hline
    $D$ [day$^{-1}$] & 0.1-0.3 \\
    $Q_d$ [L/min] & 75 \\
    $Q_h$ [L/min] & 75 \\
    $Q_{air}$ [L/min] & 250 \\
    $Q_{CO_2}$ [L/min] & 5 \\
    \hline
\end{tabular}
\caption{Value of the parameters of the standard operation.}\label{tab:tradicional}
\end{table}

\subsection{Cost function and constraints}\label{subsec:fdc}

With the approach presented in this work, the cost function $J$ must capture all the costs and revenue that come from or are affected by the operation of the reactor. Fixed outlays such as the construction or installation of the pond, as well as expenses that do not depend on operational decisions, e.g., nutrient addition, are excluded because the optimizer cannot influence them. Under these premises, the dominant operating costs are the energy expenditures of the plant, while revenues are derived from the biomass harvested.

The two most significant energy costs are: (i) the paddle wheel, which propels the culture medium, and (ii) the air blower, whose purpose is to strip dissolved oxygen from the medium. The cost of the paddle wheel scales with the square of the depth of the water: The higher the level, the more power is required. In contrast, the energy use of the blower is directly proportional to the airflow rate. A further penalty comes from downstream biomass processing, which effectively reduces the net income per gram of harvested biomass. For simplicity, the unit price of electricity is assumed constant throughout the campaign, although time‑of‑day pricing could, in practice, influence the optimal strategy.

The remaining variable costs of the process are primarily associated with raw materials, namely nutrients and carbon dioxide. Nutrient demand is proportional to the total biomass present in the culture and is supplied through the dilution water. Plant operators assess the culture condition daily and offline, then add the required nutrients to the dilution water accordingly.

Carbon dioxide, on the other hand, can be considered either a cost or a revenue stream depending on its source: it incurs a cost when using pure $\mathrm{CO_2}$, but may be treated as an income if recovered from another industrial activity, thereby partially mitigating its environmental impact. In the simulated facility, pure $\mathrm{CO_2}$ is used, and thus it is treated as a cost.

The remaining variable costs of the process are primarily associated with raw materials, mainly nutrients and carbon dioxide. Nutrient requirements are proportional to the total biomass in the medium and are supplied through the dilution water. Plant operators assess the culture status daily and add the necessary nutrients to the incoming water accordingly. On the other hand, carbon dioxide can either represent a cost when using pure $\mathrm{CO_2}$, or a potential revenue stream if sourced from another industrial activity, thereby partially offsetting its environmental impact. In the simulated facility considered in this study, pure $\mathrm{CO_2}$ is used, and therefore it is treated strictly as a cost.

Revenue can be framed in two alternative ways. An option is to define income as the total biomass \textit{produced} in the reactor over the horizon (Equation \eqref{eq:biomasacrecimiento}). This assumes that every gram eventually gets harvested and focuses solely on maximizing growth. The second option is to base the income on the total biomass actually \textit{harvested} (Equation \eqref{eq:biomasacosechado}). This definition places greater weight on the timing of harvest operations, but if not all constraints are properly handled, it may drive the system toward premature depletion, rendering the process unsustainable. This issue will be examined in depth later.

\begin{equation}\label{eq:biomasacrecimiento}
    Q_{biomass}(t)=X(t)\cdot\mu(t)\cdot V(t)
\end{equation}

\begin{equation}\label{eq:biomasacosechado}
    Q_{biomass}(t)=X(t)\cdot Q_h(t)
\end{equation}

In this work, the second approach was adopted as it is more closely aligned with the theoretical optimum by treating the harvest timing as a key factor in the cost function. Consequently, the complete cost function is defined by Equation \eqref{eq:Jcompleta}, where $\hat{P}_{biomass}$ is the net selling price of biomass (considering downstream processing costs), $C_{power}$ the cost of electric power, $P_{blower}$ the blower power consumption per bubbled unit of volume, $P_{wheel}$ the power consumption of the paddle wheel, $V_l$ the linear speed of the water, $C_{nut}$ the nutrient cost per biomass kilogram, and $C_{CO_2}$ the cost of $\mathrm{CO_2}$ per unit of volume. It is important to note that the biomass selling price used to compute the average cost was set at an intermediate value, acknowledging that it can range widely, from €1 to €100 per kilogram, depending on its intended application and potential market fluctuations. Table \ref{tab:costes} includes the value of these parameters.

\begin{align}
    J(t) =\; & -\hat{P}_{\text{biomass}}\cdot Q_{\text{biomass}}(t)
    + C_{\text{power}}\cdot P_{\text{blower}}\cdot Q_{\text{air}}(t)\dots\notag \\
    & + C_{\text{power}}\cdot P_{\text{wheel}}\cdot V_l\cdot h(t)^2 
    + C_{\text{nut}}\cdot Q_d(t) + C_{\text{CO}_2}\cdot Q_{\text{CO}_2}(t)
    \label{eq:Jcompleta}
\end{align}

\begin{table}[ht]
\centering
\begin{tabular}{r c}
    \hline
    Parameter & Value \\
    \hline
    $\hat{P}_{biomass}$ [€/kg] & 9.7 \\
    $C_{power}$ [€/kWh] & 0.086 \\
    $P_{blower}$ [kWh/m$^3$] & 0.0178 \\
    $P_{wheel}$ [kWh/m$^3$] & 0.1737 \\
    $V_l$ [m/s] & 0.2 \\
    $C_{nut}$ [€/m$^3$] & 0.3125 \\
    $C_{CO_2}$ [€/m$^3$] & 0.44 \\
    \hline
\end{tabular}
\caption{Value of the parameters of the cost function.}\label{tab:costes}
\end{table}

However, certain simplifications can be made to the cost function that significantly ease the resolution of the optimization problem. These simplifications are related to raw material costs. First, the costs associated with nutrients can be effectively embedded into the net selling price of the biomass. While this is an approximation, it is reasonable, since over the course of a day the volume of harvested water is practically equal to the volume added through dilution (aside from minor adjustments due to evaporation). Additionally, nutrient dosing is manually and periodically adjusted by the plant operators based on the conditions of the culture, specifically, its biomass concentration. This allows for the removal of an entire term from the cost function, simplifying the optimization process.

The second simplification involves removing the term associated with carbon dioxide. The carbonate balance in the model represents the most complex component, and including this term greatly increases the difficulty of finding feasible solutions. Moreover, the cost of carbon dioxide is relatively minor compared to other system costs, and the daily $\mathrm{CO_2}$ demand of the culture does not fluctuate enough to cause a significant impact on the cost function. With these simplifications, the new cost function is represented in Equation \eqref{eq:J}.

\begin{equation}
    J(t)=-\hat{P}_{biomass}\cdot Q_{biomass}(t)+C_{power}\cdot P_{blower}\cdot Q_{air}(t)+C_{power}\cdot P_{wheel}\cdot V_l\cdot h(t)^2
    \label{eq:J}
\end{equation}

Because the process variables interact in complex, nonlinear ways, the resulting optimization problem becomes non‑convex. Multiple local minima can complicate the search for a true optimum, making the choice of permissible ranges, and, critically, the quality of the initial guesses, decisive for successful convergence.

Once the cost function has been defined, it is essential to establish the constraints that will govern the optimization problem. Naturally, the primary constraints are associated with the manipulable variables of the system, namely, the operational flow rates: air flow, $\mathrm{CO_2}$ injection, dilution, and harvesting. Each of these variables is bounded within the range specified in Table \ref{tab:qmaxqmin}.

\begin{table}[ht]
\centering
\begin{tabular}{r c c}
    \hline
    Variable & Min value & Max value \\
    \hline
    $Q_{air}$ [L/min] & 0 & 500 \\
    $Q_{CO_2}$ [L/min] & 0 & 15 \\
    $Q_d$ [L/min] & 0 & 75 \\
    $Q_h$ [L/min] & 0 & 75 \\
    $h$ [cm] & 10 & - \\
    $X$ [g/L] & $X(t_0)$ & - \\
    \hline
\end{tabular}
\caption{Constraints of the optimization problem.}\label{tab:qmaxqmin}
\end{table}

The remaining constraints in the problem are related to the state variables of the system. Specifically, two states are critical to ensure continuity of operation: the reactor water level and biomass concentration. If the cost function rewards biomass harvesting, leaving these states unconstrained can lead to full system depletion at the end of each day. In such cases, the optimal strategy may be to harvest all available biomass just before shutdown, especially during non-illuminated hours, when no further growth is expected. However, this behavior undermines the sustainability of the process. Since the reactor operates continuously and is not reinoculated daily, starting each day with near-zero biomass levels is neither realistic nor desirable.

To prevent this, a lower bound was imposed on the reactor level, ensuring that it never falls below 10 cm. This not only prevents reactor drainage, but also guarantees a minimum water depth that allows the paddle wheel to circulate the medium effectively at all times.

In addition, a terminal constraint was applied to the biomass concentration, $X_b(k+N_p|k)$. This means that the constraint is only enforced at the end of the prediction horizon. This design allows for temporary violations during the optimization window, provided that growth over time restores biomass to a sustainable level. The minimum terminal value was set to be the initial condition used in all simulations. Considering that the optimizer is only executed during daylight hours (when biomass growth occurs) and that the prediction horizon never spans days, it is expected that the biomass concentration converges toward this threshold by the end of each operational cycle.

Both constraints were implemented as soft constraints, anticipating that the optimal values of the corresponding variables might approach the lower bounds of their admissible ranges, and are also presented in Table \ref{tab:qmaxqmin}. This design choice helps prevent convergence issues that could arise from modeling errors or unforeseen phenomena when a state variable slightly falls below its limit. To achieve this, slack variables were introduced, and each was assigned a large negative weight in the cost function. Taking all these elements into account, the complete optimization problem and its set of constraints are defined by Equation \eqref{eq:optimizacion}.

\begin{equation}
\begin{aligned}
    \min_{Q} \quad & \sum_{i=0}^{N_p} \left( J(k+i \mid k) + W_{S_h}\cdot S_h(k+i \mid k) \right)^2 + W_{S_{X_b}}\cdot S_{X_b}^2 \\
    \text{S.T.} \quad 
    & \hat{x}(k+i \mid k) = f\left( \hat{x}(k+i-1 \mid k), Q(k+i-1 \mid k), d(k+i-1 \mid k) \right), \\
    & Q_{\text{min}} \leq Q(k+i \mid k) \leq Q_{\text{max}}, \\
    & h_{\text{min}} \leq h(k+i \mid k) - S_h(k+i \mid k), \\
    & X_{b,\text{min}} \leq X_b(k+N_p \mid k) - S_{X_b}, \\
    & -1 \leq S_h(k+i \mid k) \leq 0, \\
    & -0.3 \leq S_{X_b} \leq 0,
\end{aligned}
\label{eq:optimizacion}
\end{equation}

In this equation, $Q$ represents the set of manipulable system variables. $S_h$ and $S_{X_b}$ are the slack variables associated with the water level and biomass concentration of the reactor, respectively. $W_{S_h}$ and $W_{S_{X_b}}$ are the weights associated to these slack variables, which will be notably high. The set $d$ includes the external disturbances that affect the system. $Q_{\text{min}}$ and $Q_{\text{max}}$ define the admissible bounds for manipulable variables, $h_{\text{min}}$ is the minimum allowable water level, and $X_{b,\text{min}}$ is the terminal lower bound for biomass concentration. The slack variables were also restricted to minimum values of - 1 cm for the water level and –0.3 g/L for the biomass concentration, respectively.

With this, the optimization problem is fully defined. Subsequent sections will detail the implementation of the optimizer, along with key parameters and specific considerations taken into account during its development.

\subsection{EMPC considerations}\label{subsec:tuning}

The core principles of EMPC were introduced earlier in Section \ref{subsec:fdc}, but several key implementation decisions require further clarification.

First, the optimizer is executed every 5 minutes. This interval is fast enough to enable effective decision-making over the manipulable variables, whose influence on the system is relatively immediate, while also being slow enough to remain compatible with the computational demands of solving the optimization problem.

The optimizer is triggered only when the global irradiance measured by the sensor exceeds 100 W/m$^2$. Below this threshold, biomass growth is minimal, and continued operation of the reactor yields little benefit. Similarly, once irradiance drops below this value at the end of the day, the optimizer is deactivated and all manipulable variables are set to zero.

The complexity of the optimization problem is determined jointly by the sampling time and the prediction horizon. Initially, the strategy is evaluated using a prediction horizon that spans the entire remaining daylight period; that is, from the moment the optimizer is activated until it is turned off at sunset. With each new optimization step, the horizon shrinks accordingly, covering only the remaining operating window. This strategy, known as a \textit{shrinking horizon}, reduces computational burden without compromising optimization quality, since the plant is not operated during nighttime hours.

This setup allows the optimizer to evaluate the long-term impact of each decision, bringing the solution closer to the theoretical optimum. However, the computational load, particularly during early hours of operation, is considerable. Therefore, it is necessary to assess whether shorter prediction horizons can achieve similarly effective results.

With this approach, the optimizer never considers decisions beyond the operation of the current day. This highlights the need for terminal constraints on biomass concentration, as discussed in Section \ref{subsec:fdc}, to prevent system depletion at the end of each cycle.

A similar trade-off exists for the control horizon. A longer control horizon typically yields results closer to the theoretical optimum but increases computational complexity and solver time. That said, the sensitivity to this parameter is generally minor. Several studies have shown that reduced control horizons can still produce near-optimal outcomes, even when they do not span the full prediction horizon \cite{rossiter2017model}.

The optimization problem is highly non-convex, which implies the presence of multiple local minima. This makes the choice of initial guesses for the control variables especially critical. However, selecting appropriate initial values is not straightforward. To address this, a default value of 5 L/min was assigned to each flow variable as the starting point for the optimization.

For the base case, it is assumed that perfect knowledge of future disturbances is available, sampled every 5 minutes. In practice, this may not hold, either due to inaccuracies in forecasts or the complete absence of predictive data. Strategies for dealing with these limitations are discussed in detail in Section \ref{subsec:perturbaciones}.

Finally, the proposed control strategy is implemented and evaluated in \textit{MATLAB}, using \textit{CasADi} \cite{andersson2019casadi}. The optimization problem is solved using \textit{IPOPT}, another open-source tool based on an interior-point method for constrained nonlinear optimization \cite{biegler2009large}. While \textit{IPOPT} does not guarantee convergence to a global optimum in non-convex settings, it is effective for a wide range of problems. In cases where local minima pose challenges, the optimizer could be combined with genetic algorithms or other global search methods to improve the quality of the initial guess.

\subsection{EMPC results}\label{subsec:resultados}

The proposed strategy will be evaluated both qualitatively and quantitatively across different scenarios and compared against the traditional operation described in Section \ref{subsec:tradicional}. Given the extensive availability of climatic data from the previously described facilities across various times of the year, the system can be simulated under a wide range of environmental conditions.

\begin{figure}[!ht]
    \centering
    \includegraphics[width=0.9\textwidth]{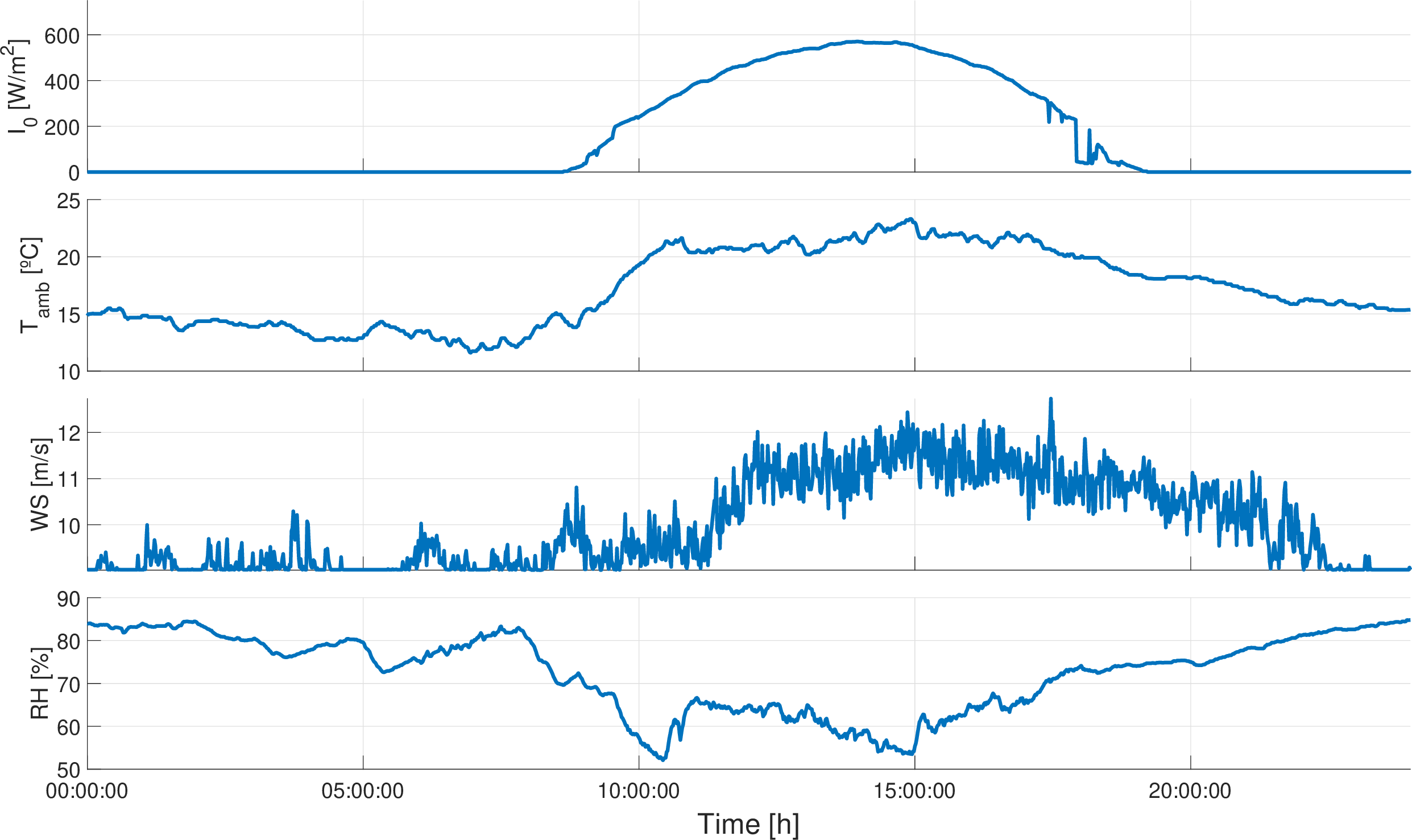}
    \caption{Environmental conditions for October 28th, 2023. From top to bottom, the profiles of global irradiance, ambient temperature, wind speed, and relative humidity.}
    \label{fig:perfiles}
\end{figure}

Before comparing the proposed approach with the traditional operating strategy, it is useful to assess the influence of the prediction horizon on the quality of the solution. To this end, one representative day was selected as benchmark to determine whether near-optimal solutions can be achieved with shorter prediction horizons. Reducing this parameter can significantly lower the computation time of the optimizer, which becomes especially relevant when evaluating the strategy over a large number of days. The disturbance profiles for the selected day are shown in Figure \ref{fig:perfiles}.

This scenario is used to evaluate the potential benefits of employing a long prediction horizon that spans until the end of the day. Such an approach theoretically yields the optimal solution, but it also greatly increases the complexity of the optimization problem. This translates into longer computation times and, in some cases, the inability to reach a global optimum before the optimizer hits its iteration limit.

An alternative is to use shorter prediction horizons. Specifically, a 2-hour horizon (12 samples) was selected, as it is sufficient to capture the settling times of the key dynamic variables of the system. When the end of the operation approaches, the \textit{receding horizon} is replaced by a \textit{shrinking horizon} approach.

\begin{figure}[!ht]
    \centering
    \begin{subfigure}[b]{0.85\textwidth}
        \centering
        \includegraphics[width=\textwidth]{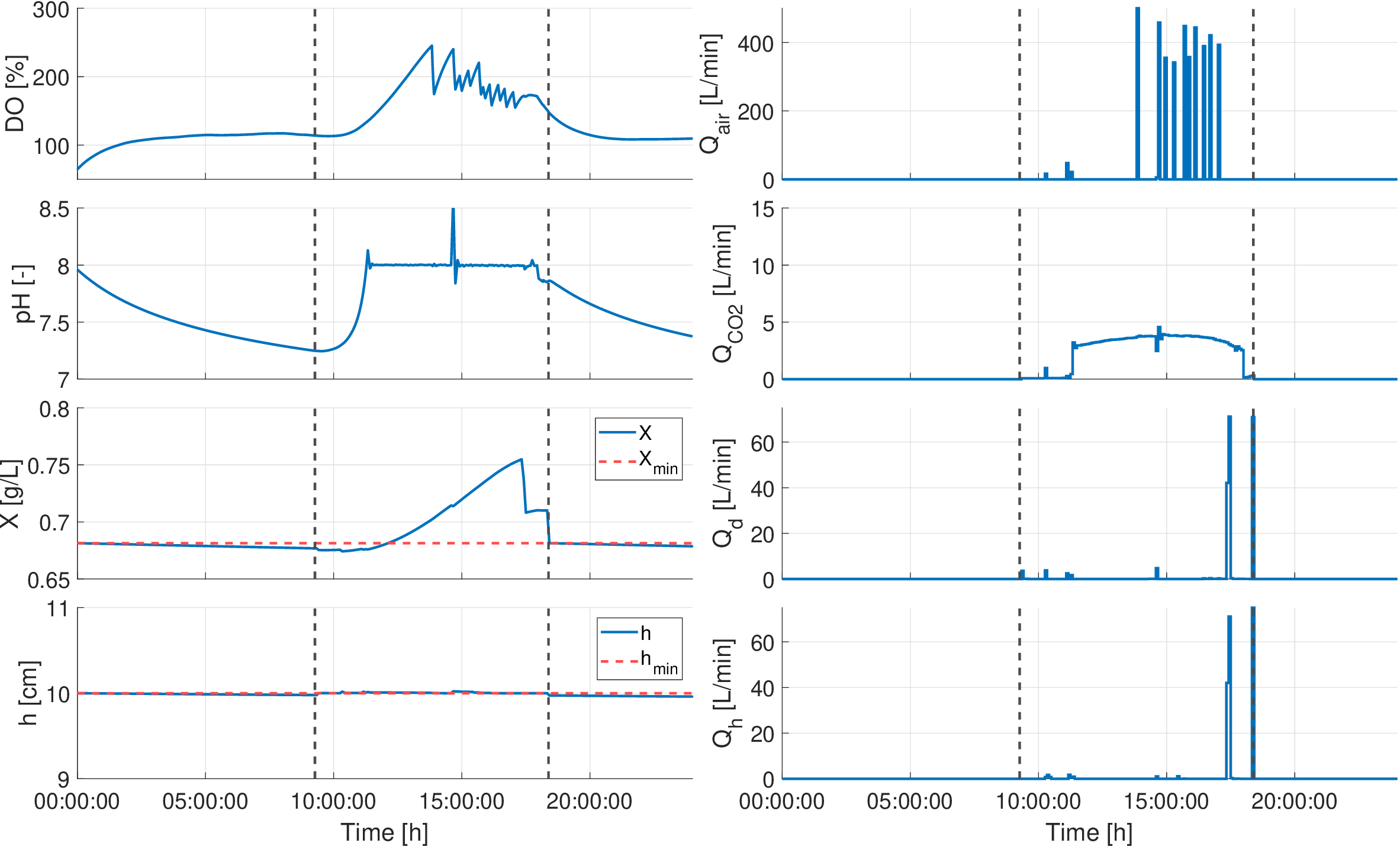}
        \caption{Optimization results with a prediction horizon covering the entire day. Cost function value of -2.3298€.}
        \label{fig:resultadoshorizonte24h}
    \end{subfigure}
    \vspace{0.5cm}
    \begin{subfigure}[b]{0.85\textwidth}
        \centering
        \includegraphics[width=\textwidth]{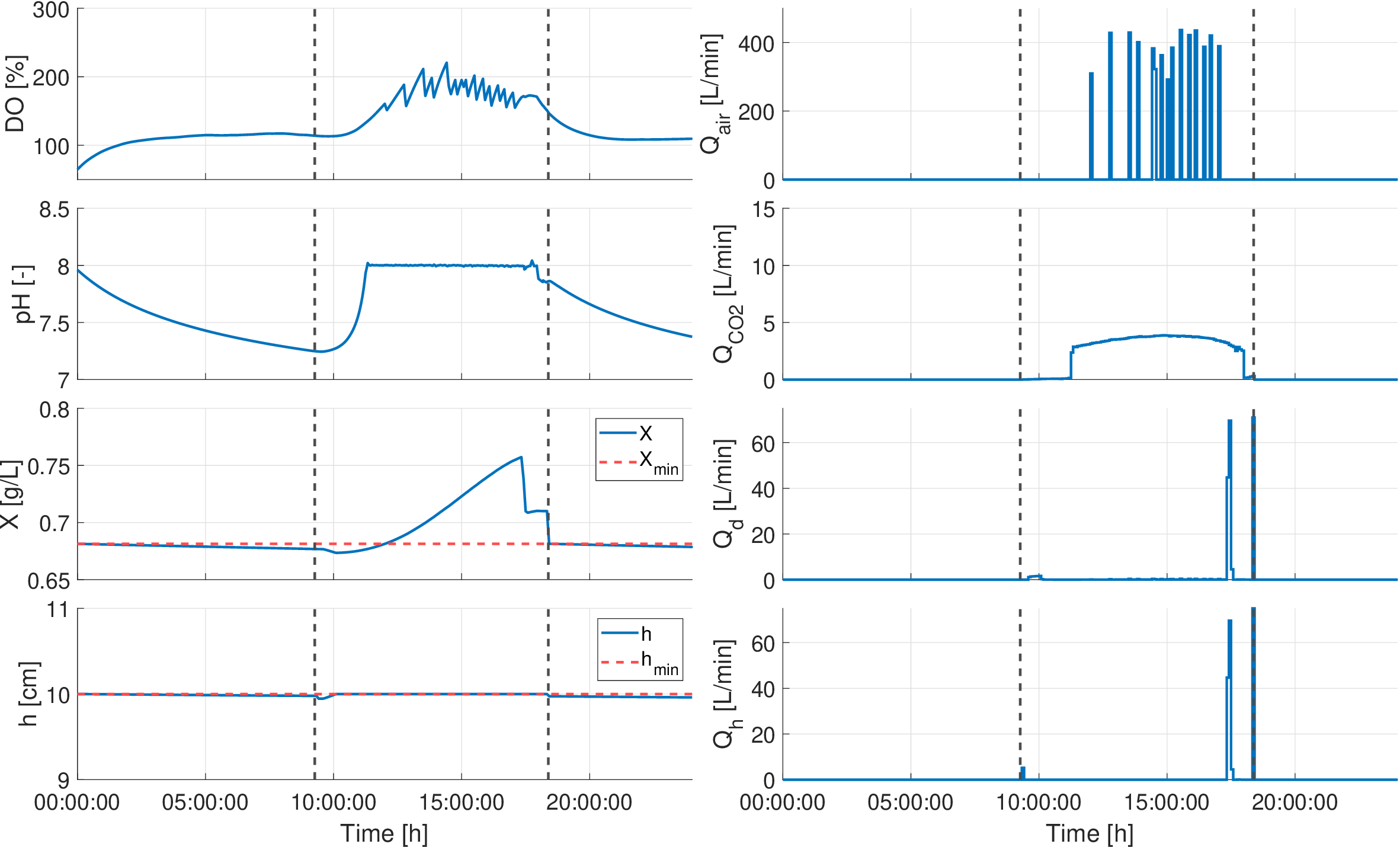}
        \caption{Optimization results with a prediction horizon of 2 hours. Cost function value of -2.3759€.}
        \label{fig:resultadoshorizonte2h}
    \end{subfigure}
    \caption{Comparison between a full-day and a 2-hour prediction horizon for October 28th, 2023. In the left, from top to bottom, the profiles of dissolved oxygen, pH, biomass concentration and water level. In the right, the air, $\mathrm{CO_2}$, dilution and harvesting flow rates.}
    \label{fig:comparacionhorizontes}
\end{figure}

Figure \ref{fig:comparacionhorizontes} compares both prediction horizons for a full day of operation. The left panel shows the main system states, which are the most influential on the cost function, in the following order: dissolved oxygen, pH, biomass concentration, and water level. The right panel displays the manipulable variables, in order: air flow rate, $\mathrm{CO_2}$ flow rate, dilution flow rate, and harvesting flow rate. The vertical dashed black lines represent the beginning and the end of the operation, while the dashed red lines represent the lower biomass concentration and water level constraints.

Without yet delving into the reasoning behind the output of the optimizer, it is clear that both strategies pursue similar operational goals. In fact, the optimizer using the longer prediction horizon exhibits less stable pH control and introduces minor dilution pulses throughout the day, whereas the 2-hour horizon alternative results in smoother, more consistent profiles. This difference is mainly due to the increased complexity of the full-horizon problem, which in some optimization steps prevents the solver from converging to a global optimum within the imposed iteration limit.

In terms of computation time, the optimizer with the long prediction horizon required over 57 hours to complete the simulation, in contrast to approximately 2 hours and 40 minutes for the shorter-horizon case. More importantly, the final cost function value reached with the long-horizon optimizer was –2.3298€, compared to –2.3759€ obtained with the simpler alternative. It may seem counterintuitive that the more complex configuration performs worse, but again, this is attributed to its inability to find the global optimum within the allotted time.

This limitation could be overcome by allowing the optimizer more time to converge. However, considering how similar the results are between both strategies, it becomes evident that the optimal solution lies close to the one achieved with the 2-hour horizon. For this reason, the simpler and more computationally efficient approach was adopted.

Examining the optimal control behavior proposed by the optimizer, one can observe that the pH is driven to and maintained at a constant value of 8 throughout the operation. This is entirely reasonable, as the model gives that a pH of 8 corresponds to the optimal growth conditions for the microalgae, and the use of $\mathrm{CO_2}$ for pH regulation incurs no economic cost. Furthermore, the main control input influencing pH, the $\mathrm{CO_2}$ flow rate, affects virtually no other variable in the system.

This observation opens the door to a meaningful simplification of the optimization strategy. Instead of explicitly including pH regulation in the EMPC problem, pH control can be delegated to a separate feedback controller (e.g., a PID), tasked with maintaining it near the optimal value of 8.

From the perspective of the optimizer, pH can be assumed to remain at 8 throughout the entire prediction horizon. While this assumption introduces a slight inaccuracy during the early hours of the day, when the optimizer might overestimate the expected growth rate due to suboptimal pH conditions, these early hours typically involve passive operation, so the effect on overall performance is negligible. For the remainder of the day, assuming pH = 8 is valid, provided the feedback controller is well-tuned. Minor deviations from this value will have minimal impact on growth.

This modification allows for the removal of three state variables from the first-principles model used by the optimizer: the concentrations of carbon dioxide, total inorganic carbon (TIC), and hydrogen ions. It also eliminates one control variable from the decision set. The resulting simplification significantly reduces the complexity of the optimization problem, as these states are highly interconnected with most of the manipulable variables, disturbances, and internal dynamics of the system.

A PI controller with anti-windup action (see Equation \eqref{eq:PI}) was selected for pH regulation, based on the fact that pH exhibits first-order dynamics with respect to the $\mathrm{CO_2}$ flow rate, as described in Equation \eqref{eq:funcionprimerorden}. The controller was tuned using the \textit{SIMC} method, which requires only a first-order model of the system. Given the highly nonlinear nature of pH dynamics, a nominal model with average parameters was used for tuning. Table \ref{tab:parametrosmodelo} shows the typical ranges for the model parameters, along with their corresponding midpoints \cite{caparroz2024novel}.

\begin{equation}\label{eq:PI}
    C_{PI}(s)=K_p\cdot\left(1+\frac{1}{T_i\cdot s}\right)
\end{equation}

\begin{equation}\label{eq:funcionprimerorden}
    G(s) =\frac{pH(s)}{Q_{CO_2}(s)}=\frac{k}{\tau\cdot s+1}e^{-L\cdot s}
\end{equation}

\begin{table}[ht]
\centering
\begin{tabular}{r c c c}
    \hline
    Parameter & Min value & Max value & Mean value \\
    \hline
    $k$ [min/L] & -1 & -0.1 & -0.55 \\
    $\tau$ [s] & 500 & 2000 & 1250 \\
    $L$ [s] & 300 & 300 & 300 \\
    \hline
\end{tabular}
\caption{Transfer function parameter ranges and mean values.}\label{tab:parametrosmodelo}
\end{table}

The SIMC method uses the desired closed-loop time constant of the system $\tau_c$ as the main tuning parameter for the controller \cite{skogestad2003simple}. For a relatively slow system, a common and effective choice is to set this time constant equal to the time delay of the system, which in this case is fixed at 300 seconds. With this configuration, Table \ref{tab:parametroscontrolador} presents the resulting controller parameter values.

\begin{table}[ht]
\centering
\begin{tabular}{r c c c}
    \hline
    Parameter & Tuning rule & Value \\
    \hline
    $K_p$ [L/min] & $\frac{1}{k}\frac{\tau}{\tau_c+L}$ & -3.8 \\
    $T_i$ [s] & $\min\{\tau, 4(\tau_c+L)\}$ & 1250 \\
    \hline
\end{tabular}
\caption{Controller parameter values applying SIMC tuning rules \cite{skogestad2003simple}.}\label{tab:parametroscontrolador}
\end{table}

The controller was enhanced with anti-windup protection to prevent saturation issues. A back-calculation scheme was implemented, using a tracking constant of $T_t = 1$ seconds, working as a clipping action at the beginning of the day, prioritizing keeping the integral action low during prolonged saturation periods. Once pH has reached 8 for the first time in the day, this parameter can be increased to equal $T_t=T_i=1250$ seconds, although at this point there should be no problem with input saturation.

\begin{figure}[!ht]
    \centering
    \includegraphics[width=\textwidth]{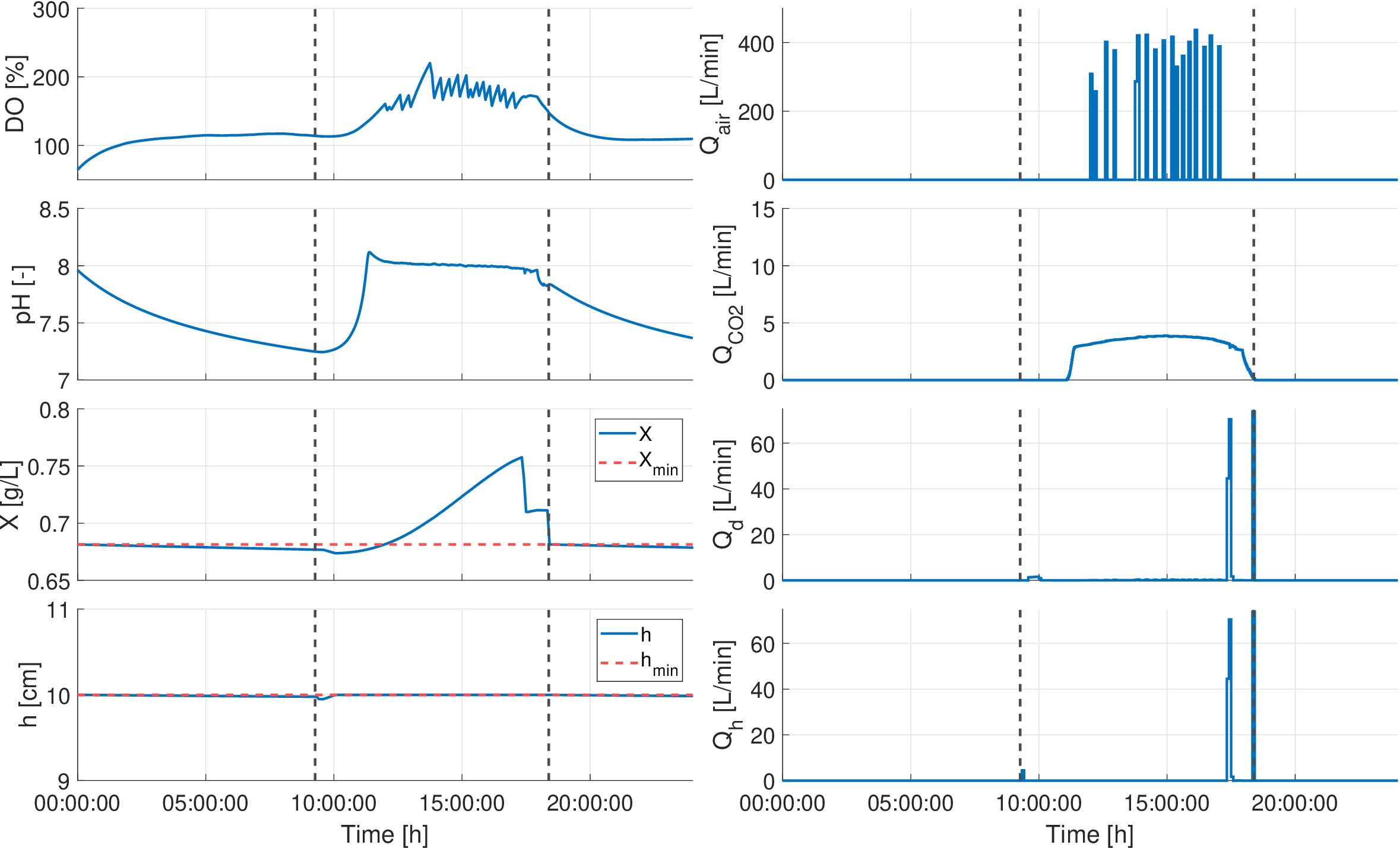}
    \caption{Optimization results combined with pH PI control for October 28th, 2023. Cost function value of -2.2750€. In the left, from top to bottom, the profiles of dissolved oxygen, pH, biomass concentration and water level. In the right, the air, $\mathrm{CO_2}$, dilution and harvesting flow rates.}
    \label{fig:resultadosPIpH}
\end{figure}

The main challenge faced by this controller is the rapid rise in pH during the early hours of operation. This rise develops significant momentum, which is difficult to anticipate with a non-predictive controller. As a result, the system exhibits overshoot in the pH value, from which the controller takes several minutes to recover. To mitigate this behavior, the proportional gain was significantly increased to $K_p = -7.5$ [L/min]. Figure \ref{fig:resultadosPIpH} shows the results of the same test previously performed, now evaluated with this updated control approach.

In this case, the cost function reached a final value of –2.2750€. Although this result is slightly less favorable than those obtained with previous approaches, it is strongly justified by the about 10 times reduction in total simulation time, which dropped to under 15 minutes. This same approach can be tested using a full-day control horizon. However, while this slightly improves the final cost function value (-2.2812€), the computation time increases significantly, reaching 113 minutes. Table \ref{tab:resultadoshorizontes} summarizes the results obtained for each approach, along with their respective computation times.

\begin{table}[!ht]
\centering
\begin{tabular}{r c c c}
    \hline
    Prediction horizon & Optimizing pH & Cost function & Elapsed time \\
    \hline
    Full-day & Yes & -2.3298 & 57 hours \\
    2 hour & Yes & -2.3759 & 160 min \\
    Full-day & No & -2.2812 & 113 min \\
    2 hour & No & -2.2750 & 15 min \\
    \hline
\end{tabular}
\caption{Cost function value and elapsed time for each of the proposed approaches.}\label{tab:resultadoshorizontes}
\end{table}

Based on this information, the simpler approach was selected for the subsequent simulations, as the improvement in the cost function was not considered significant enough to justify the substantially longer computation times. This makes the approach much more suitable for long simulation campaigns, which are essential for assessing the potential advantages of optimized operation over traditional strategies, and for real-time implementation in physical systems. Table \ref{tab:parametrosEMPC} summarizes the final value of the different EMPC tuning parameters.

\begin{table}[!ht]
\centering
\begin{tabular}{r c}
    \hline
    Parameter & Value \\
    \hline
    $T_s$ [min] & 5 \\
    $N_p$ [samples] & 24 \\
    $N_c$ [samples] & 6 \\
    $W_{S_h}$ [€$\cdot$cm$^{-2}$/s] & 10$^4$ \\
    $W_{S_{X_b}}$ [€$\cdot$L$^2\cdot$g$^{-2}$/s] & 10$^{11}$ \\
    $S_{h,min}$ [cm] & -1 \\
    $S_{X_b,min}$ [g/L] & -0.3\\
    \hline
\end{tabular}
\caption{Final value of EMPC tuning parameters.}\label{tab:parametrosEMPC}
\end{table}

\begin{figure}[!ht]
    \centering
    \begin{subfigure}[b]{0.9\textwidth}
        \centering
        \includegraphics[width=\textwidth]{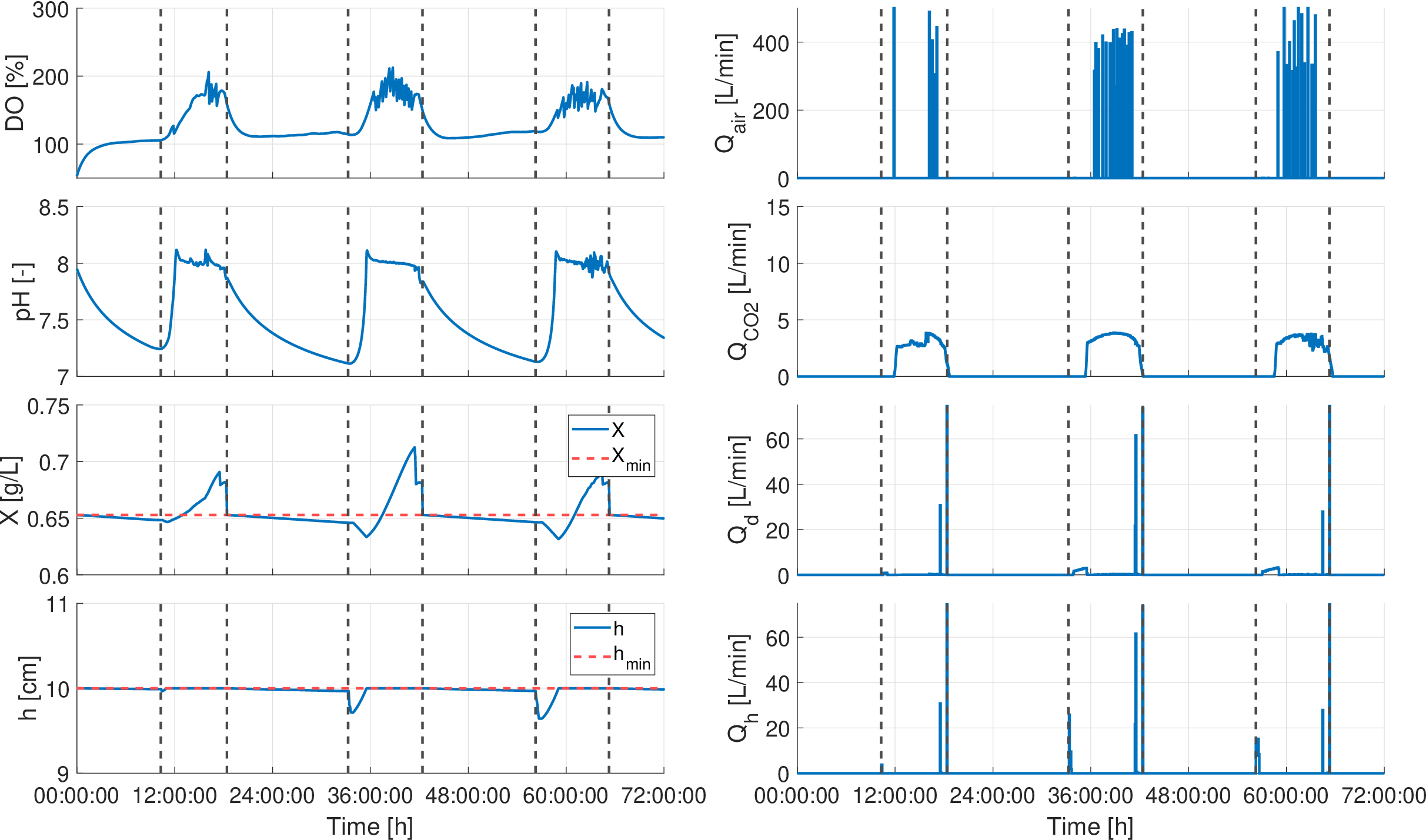}
        \caption{Three-day optimized operation. Cost function value of -3.0227€.}
        \label{fig:resultadosoptimizadoroctubre}
    \end{subfigure}
    \vspace{0.5cm}
    \begin{subfigure}[b]{0.9\textwidth}
        \centering
        \includegraphics[width=\textwidth]{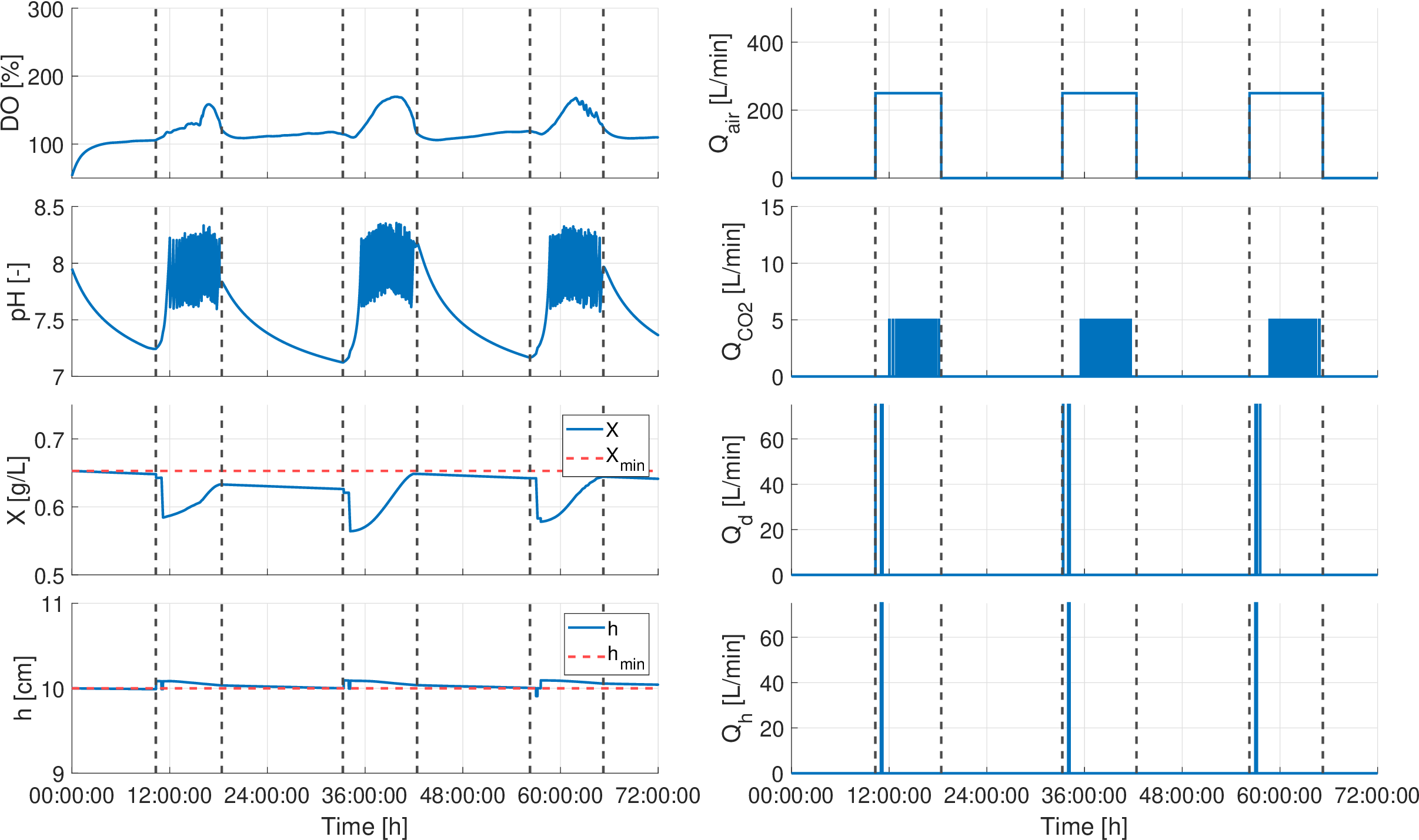}
        \caption{Three-day standard operation. Cost function value of -2.1823€.}
        \label{fig:resultadossimpleoctubre}
    \end{subfigure}
    \caption{Comparison between optimized and standard operation for a three-day autumn period (October 27th-October 29th, 2023). In the left, from top to bottom, the profiles of dissolved oxygen, pH, biomass concentration and water level. In the right, the air, $\mathrm{CO_2}$, dilution and harvesting flow rates.}
    \label{fig:resultadosoctubre}
\end{figure}

\begin{figure}[!ht]
    \centering
    \begin{subfigure}[b]{0.9\textwidth}
        \centering
        \includegraphics[width=\textwidth]{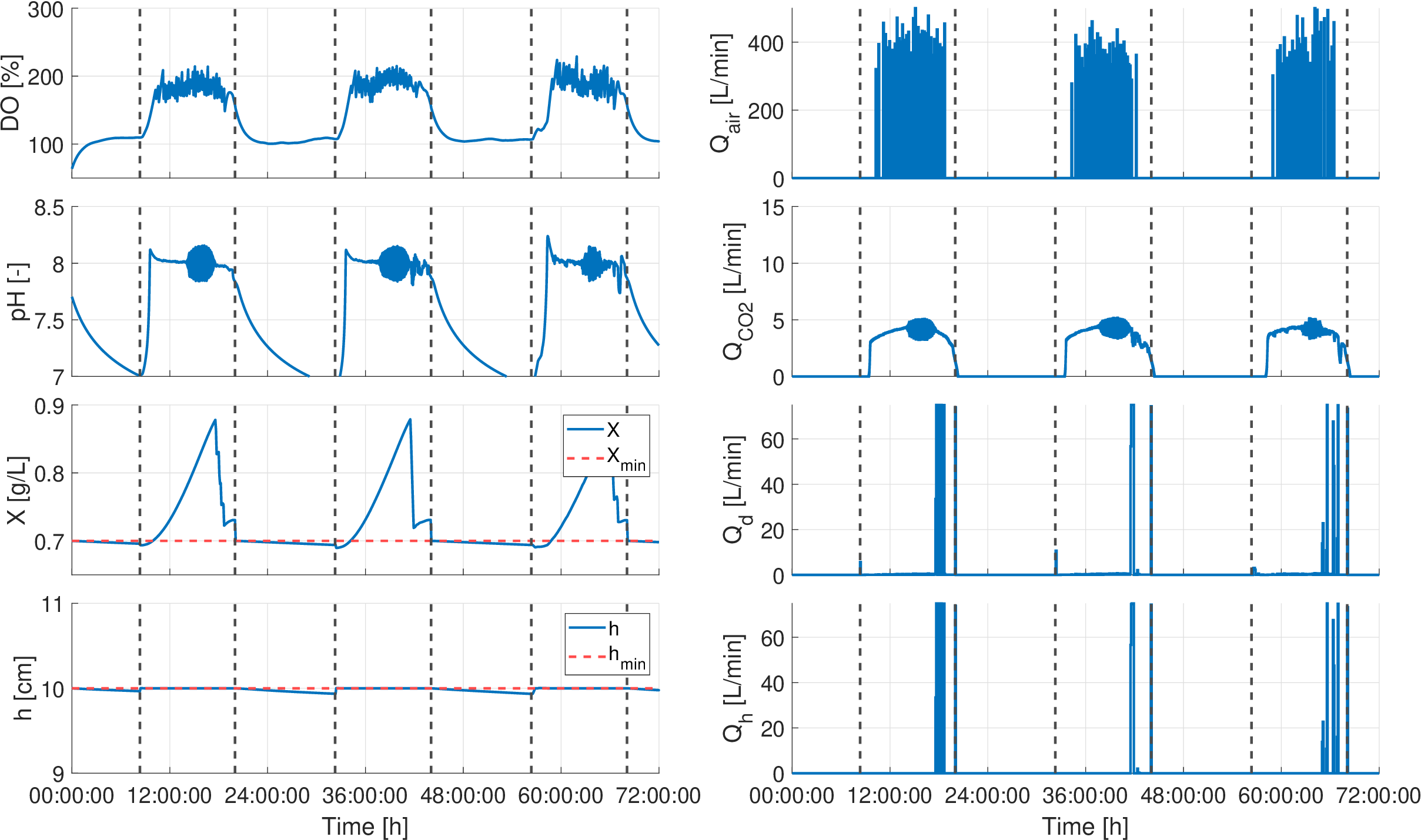}
        \caption{Three-day optimized operation. Cost function value of -26.9739€.}
        \label{fig:resultadosoptimizadorabril}
    \end{subfigure}
    \vspace{0.5cm}
    \begin{subfigure}[b]{0.9\textwidth}
        \centering
        \includegraphics[width=\textwidth]{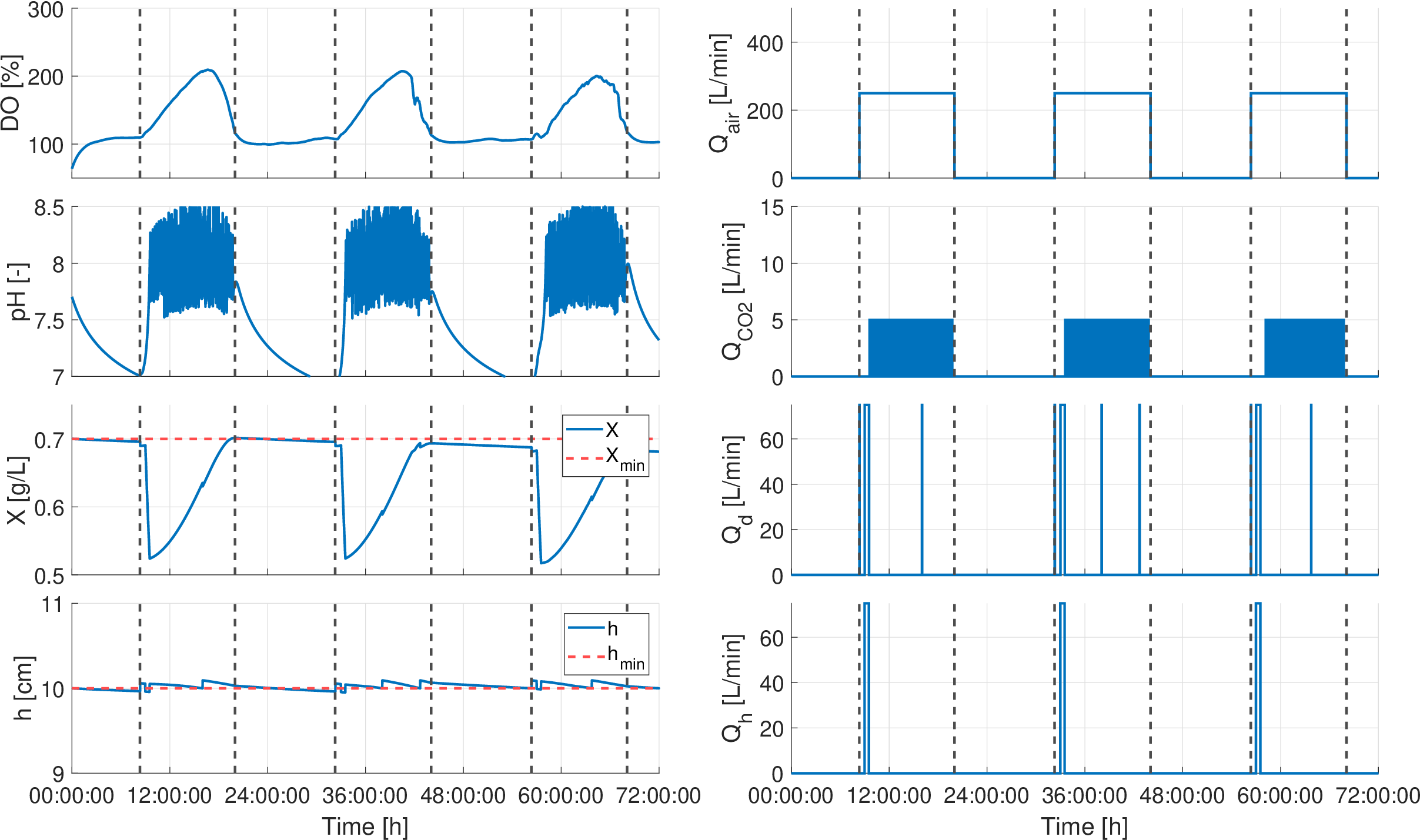}
        \caption{Three-day standard operation. Cost function value of -21.9610€.}
        \label{fig:resultadossimpleabril}
    \end{subfigure}
    \caption{Comparison between optimized and standard operation for a three-day spring period (April 23rd-April 25th, 2025). Cost function value of . In the left, from top to bottom, the profiles of dissolved oxygen, pH, biomass concentration and water level. In the right, the air, $\mathrm{CO_2}$, dilution and harvesting flow rates.}
    \label{fig:resultadosabril}
\end{figure}

At this point, it is useful to evaluate the performance of the controller graphically in contrast with the traditional operation. This comparison is conducted over an extended time frame to assess ability of each strategy to maintain sustainable operation over time. Moreover, the scenarios were selected to represent different climatic conditions, demonstrating the adaptability of each strategy. The selected scenarios correspond to three consecutive days in October 2023 and April 2025. It is important to note that the dilution volume percentage of the reactor varies with the time of year; 10\% in October and 30\% in April. Additionally, the PI controller gain was adjusted between scenarios to better fit system dynamics. Simulation results for each strategy under these scenarios are shown in Figures \ref{fig:resultadosoctubre} and \ref{fig:resultadosabril}.

The results can be analyzed both quantitatively and qualitatively. Quantitatively, in October, the optimized operation yields a final cost function value of –3.0227€, compared to –2.1823€ for the traditional approach. The economic gains during this period are quite limited due to unfavorable weather conditions, which result in a low growth rate of the culture. In April, the values differ even more significantly: –26.9739€ for the optimized operation and –21.9610€ for the traditional one. The variation in profitability across seasons is notable and understandable, as the main revenue source of the system is harvested biomass, which is directly linked to growth rates, largely driven by incident solar radiation. During operational hours, the average irradiance received in October is 355 W/m$^2$, while in April it increases to 530 W/m$^2$.

These results clearly demonstrate that the optimized strategy outperforms the traditional approach in every scenario. It not only delivers higher economic returns, but also adapts more effectively to daily fluctuations, leading to a more resilient and sustainable operation.

A key advantage of the optimized strategy is its ability to consistently meet the terminal biomass concentration constraint, ensuring that the operation of each day is self-contained and independent from previous days. In contrast, the traditional approach lacks this safeguard, allowing biomass concentration to drop below initial levels by the end of the day. If this occurs over several consecutive days, system performance can deteriorate severely, potentially requiring reactor reinoculation to restore proper operation.

The results also highlight the strong seasonal dependency of the process. In April, when growth conditions are more favorable, profit margins are considerably higher. This underscores the importance of maintaining optimal operation during peak productivity periods, as these months have a disproportionate impact on the overall economic viability of the system.

Qualitatively, the optimized strategy reveals several consistent operational policies. Dissolved oxygen is tightly regulated using short, high-intensity air flow bursts. This minimizes total blower usage while keeping oxygen levels low. The culture depth is maintained at 10 cm throughout most of the operation, only occasionally dropping below this value. This helps reduce paddle wheel energy consumption and slightly increases culture temperature, bringing it closer to the optimal range for this species.

One of the most interesting outcomes relates to biomass concentration. Previous studies have shown that, at constant depth, the optimal biomass concentration varies with incident radiation \cite{otalora2024modeling}. Accordingly, during the early and late hours of the day, it is preferable to maintain lower biomass concentrations to maximize growth. On days with lower radiation, the biomass concentration is also kept lower overall.

Furthermore, the optimizer often drives biomass concentration toward its lower constraint at the end of the day, when no further growth occurs, in order to maximize revenue through harvesting. Interestingly, in most cases, this final reduction in biomass concentration occurs in two steps rather than a single large decrease. The rationale behind this is aligned with the earlier point: maintaining lower concentrations during low-radiation hours enhances growth in the final part of the day. All these effects are clearly visible in the biomass concentration profiles shown in Figures \ref{fig:resultadosoctubre} and \ref{fig:resultadosabril}.

Finally, it is also valuable to establish a quantitative comparison across a larger number of scenarios. To this end, a total of 14 different days from April 19th to May 8th of the year 2025 were selected, and the values of the cost function obtained through the optimized strategy and the traditional operation were compared. Table \ref{tab:resultadosfinales} summarizes the results of these simulations.

\begin{table}[ht]
\centering
\begin{tabular}{rccc}
    \hline
    Operation & Cost function & Best day & Worst day \\
    \hline
    Optimized & -136.2447 & -11.7947 & -4.4555 \\
    Traditional & -99.2737 & -8.5848 & -5.9523 \\
    \hline
\end{tabular}
\caption{Simulation results after a 2-week simulation scenario (in the period April 19th-May 8th, 2025).}\label{tab:resultadosfinales}
\end{table}

Several important conclusions can be drawn from these results. The most direct and significant finding is that the optimized operation outperforms the traditional approach, achieving over 33\% higher profits. This improvement stems not only from optimizing the cost function using system knowledge and disturbance forecasts, but also from effectively managing constraints to ensure smoother and more sustainable operation over time.

This latter factor also explains why the optimized strategy performs slightly worse on its worst-case day. On less sunny days, the optimizer limits the amount of biomass harvested, extracting less biomass in the short term but preserving better operating conditions for the following days instead of compromising future performance. Naturally, under favorable conditions, such as sunny days, the optimized strategy also achieves superior results, maximizing both growth and profitability.

\subsection{Disturbance prediction approaches}\label{subsec:perturbaciones}

As commented above, in the results presented in previous sections, it was assumed that perfect knowledge of future disturbances was available. However, in some cases, predictions of the disturbances affecting the system may not be available with the desired sampling time or prediction horizon. Therefore, this section proposes and compares different alternatives that can partially address this issue.

The first and closest approach to the one already proposed is to use the disturbance profile from the previous day as a forecast for the current day. This is applicable when no forecast at all is available, although it may lead to issues if weather conditions change significantly from one day to the next. 

Another alternative is to assume that disturbance forecasts are available but with a lower sampling frequency than required by the optimizer. This is the case, for example, in the real facilities on which the simulator is based, where the Weatherbit.io API provides weather forecasts up to one day ahead, sampled every hour.

The Weatherbit.io API \cite{weatherbit2025} is a web-based service that provides structured access to weather data. It offers real-time, forecast, and historical weather information, including temperature, humidity, solar radiation, wind speed, and atmospheric pressure. Designed for global coverage, the API supports various time resolutions, making it suitable for different applications.

In this case, the sample time is sufficient to resample the predicted disturbance values, which can be done using linear interpolation or splines. For simplicity, the optimizer will be evaluated assuming linearly interpolated future disturbance values. Table \ref{tab:predicciones} presents the cost function values resulting from the evaluation of each of the proposed disturbance prediction approaches, using the disturbance profiles shown in Figure \ref{fig:perfiles} as a reference.

\begin{table}[ht]
\centering
\begin{tabular}{rc}
    \hline
    Prediction approach & Cost function \\
    \hline
    Full knowledge & -2.2750 \\
    Constant values & -2.2278 \\
    Weather forecast & -2.2694 \\
    \hline
\end{tabular}
\caption{Cost function values for different disturbance prediction approaches for October, 28th 2023.}\label{tab:predicciones}
\end{table}

As expected, the approach that uses the available weather forecasts delivers results very similar to those obtained with perfect knowledge of the disturbances, as its predictions are closest to the actual conditions. In contrast, assuming constant values yields a worse performance. This is logical since, during the early hours of the day, the optimizer is unaware that radiation will increase, leading it to conclude that high biomass concentrations are never needed and causing premature dilution.

\section{Conclusions}\label{sec:conclusiones}

This paper has presented a dynamic optimization framework for the operation of raceway reactors in industrial microalgae production, focusing primarily on economic performance. Considering the biological, highly nonlinear, and time-varying nature of the process, which introduces major challenges for modeling, control, and optimization, a centralized Economic Model Predictive Control (EMPC) strategy was proposed to achieve theoretically optimal operation.

The results obtained confirm that the proposed strategy effectively ensures both economic optimization and dynamic stability of the microalgae production process. The optimized approach clearly outperforms traditional standard operation, yielding higher economic returns while adapting more efficiently to daily changes in environmental conditions. This leads to more sustainable long-term operation, with significantly wider profit margins during high-productivity periods such as April, compared to months like October with lower productivity.

Moreover, the research provides valuable insight into the key priorities and operational trade-offs that should guide decision-making in large-scale microalgae systems aiming to improve both efficiency and profitability. The optimized strategy revealed consistent and interpretable operational patterns. For example, dissolved oxygen is tightly regulated through brief but intense bursts of aeration, which reduce total air consumption while maintaining low oxygen levels. The cultivation depth is generally kept at 10 cm throughout the day, helping to minimize energy consumption from the paddle wheel and keep the culture temperature within the optimal range for the species. Biomass concentration is managed dynamically, remaining lower during low-irradiance periods, such as early and late hours of the day or cloudy conditions, to favor growth. Toward the end of the day, biomass is typically reduced to its lower bound to maximize harvest revenue, often in two steps to enhance growth during the last hours of light.

A significant improvement in computational efficiency was also achieved by assigning pH regulation to a separate feedback controller, specifically a PI controller. Although direct pH control through EMPC could offer slightly better stability, this simplification removes three state variables and one control variable from the optimization problem, drastically reducing computational complexity. Thanks to this modification, total simulation time was reduced from over 57 hours to less than 15 minutes, making the strategy much more viable for large-scale simulation campaigns and real-time implementation in physical systems.

For future work, it will be highly valuable to evaluate the performance of the proposed strategy on the actual system, addressing a more complex scenario with potential unmodeled phenomena. Additionally, insights gained from understanding the optimal operation can be used to develop simpler control policies that bring the performance of the system closer to the optimal level without the need to solve highly complex optimization problems.

Altogether, the results strongly support the adoption of optimal controllers over traditional, rule-based operation. This work represents a step forward toward more economically viable and sustainable industrial applications of microalgae cultivation.

\section*{Acknowledgments}
This work has been financed by the following projects: PID2023-150739OB-I00 project financed by the Spanish Ministry of Science and also by the European Union (Grant agreement IDs: 101060991, REALM;101146861, NIAGARA; 101214199, ALLIANCE)”). 


\bibliographystyle{elsarticle-num} 
\bibliography{bibliography}

@inbook{acien2017microalgae,
    author    = {Francisco Gabriel Acién and Jose Maria Fernández-Sevilla and Emilio Molina Grima},
    title     = {Microalgae: The Basis of Mankind Sustainability},
    booktitle = {Case Study of Innovative Projects},
    publisher = {IntechOpen},
    address   = {Rijeka},
    year      = {2017},
    chapter   = {7},
    pages     = {123--140},
    doi       = {10.5772/67930}
}

@incollection{acien2017economics,
    title     = {Economics of microalgae production},
    chapter   = {20},
    editor    = {Cristina Gonzalez-Fernandez and Raúl Muñoz},
    booktitle = {Microalgae-Based Biofuels and Bioproducts},
    publisher = {Woodhead Publishing},
    pages     = {485-503},
    year      = {2017},
    series    = {Woodhead Publishing Series in Energy},
    isbn      = {978-0-08-101023-5},
    doi       = {10.1016/B978-0-08-101023-5.00020-0},
    author    = {F.G. Acién and E. Molina and J.M. Fernández-Sevilla and M. Barbosa and L. Gouveia and C. Sepúlveda and J. Bazaes and Z. Arbib}
}

@Inbook{rahman2020food,
    author    = {Rahman, Khondokar M.},
    title     = {Food and High Value Products from Microalgae: Market Opportunities and Challenges},
    chapter   = {1},
    bookTitle = {Microalgae Biotechnology for Food, Health and High Value Products},
    year      = {2020},
    publisher = {Springer Singapore},
    address   = {Singapore},
    pages     = {3--27},
    isbn      = {978-981-15-0169-2},
    doi       = {10.1007/978-981-15-0169-2\_1}
}

@inbook{guo2020microalgae,
    author    = {Guo, Suolian and Wang, Ping and Wang, Xinlei and Zou, Meng and Liu, Chunxue and Hao, Jihong},
    title     = {Microalgae as Biofertilizer in Modern Agriculture},
    chapter   = {12},
    bookTitle = {Microalgae Biotechnology for Food, Health and High Value Products},
    year      = {2020},
    publisher = {Springer Singapore},
    address   = {Singapore},
    pages     = {397--411},
    isbn      = {978-981-15-0169-2},
    doi       = {10.1007/978-981-15-0169-2\_12}
}

@article{lam2012microalgae,
    title = {Microalgae biofuels: A critical review of issues, problems and the way forward},
    journal = {Biotechnology Advances},
    volume = {30},
    number = {3},
    pages = {673-690},
    year = {2012},
    issn = {0734-9750},
    doi = {https://doi.org/10.1016/j.biotechadv.2011.11.008},
    author = {Man Kee Lam and Keat Teong Lee}
}

@article{sanchez2021wastewater,
    title={Wastewater treatment using Scenedesmus almeriensis: effect of operational conditions on the composition of the microalgae-bacteria consortia},
    author={S{\'a}nchez-Zurano, Ana and Lafarga, Tom{\'a}s and Morales-Amaral, Mar{\'\i}a del Mar and G{\'o}mez-Serrano, Cintia and Fern{\'a}ndez-Sevilla, Jos{\'e} Mar{\'\i}a and Aci{\'e}n, Francisco Gabriel and Molina-Grima, Emilio},
    journal={Journal of Applied Phycology},
    volume={33},
    pages={3885--3897},
    year={2021},
    publisher={Springer},
    doi = {10.1007/s10811-021-02600-2}
}

@article{razzak2017biological,
    title = {Biological CO2 fixation with production of microalgae in wastewater – A review},
    journal = {Renewable and Sustainable Energy Reviews},
    volume = {76},
    pages = {379-390},
    year = {2017},
    issn = {1364-0321},
    doi = {10.1016/j.rser.2017.02.038},
    author = {Shaikh Abdur Razzak and Saad Aldin M. Ali and Mohammad Mozahar Hossain and Hugo deLasa}
}

@article{narala2016comparison,
    title={Comparison of microalgae cultivation in photobioreactor, open raceway pond, and a two-stage hybrid system},
    author={Narala, Rakesh R and Garg, Sourabh and Sharma, Kalpesh K and Thomas-Hall, Skye R and Deme, Miklos and Li, Yan and Schenk, Peer M},
    journal={Frontiers in Energy Research},
    volume={4},
    pages={29},
    year={2016},
    publisher={Frontiers Media SA},
    doi = {10.3389/fenrg.2016.00029 }
}

@article{posten2009design,
    author = {Posten, Clemens},
    title = {Design principles of photo-bioreactors for cultivation of microalgae},
    journal = {Engineering in Life Sciences},
    volume = {9},
    number = {3},
    pages = {165-177},
    doi = {10.1002/elsc.200900003},
    year = {2009}
}

@article{caparroz2024novel,
    title = {A novel data-driven model for prediction and adaptive control of {pH} in raceway reactor for microalgae cultivation},
    journal = {New Biotechnology},
    volume = {82},
    pages = {1-13},
    year = {2024},
    issn = {1871-6784},
    doi = {10.1016/j.nbt.2024.04.001},
    author = {Caparroz, M. and Guzmán, José Luis and Berenguel, M. and Acién, Francisco Gabriel}
}

@article{sanchezzurano2021abaco,
    author = {Sánchez-Zurano, Ana and Rodríguez-Miranda, Enrique and Guzmán, José Luis and Acién, Francisco Gabriel and Fernández-Sevilla, José M. and Molina Grima, Emilio},
    title = {{ABACO}: A New Model of Microalgae-Bacteria Consortia for Biological Treatment of Wastewaters},
    journal = {Applied Sciences},
    volume = {11},
    year = {2021},
    number = {3},
    article-number = {998},
    issn = {2076-3417},
    doi = {10.3390/app11030998}
}

@Inbook{bernard2016modelling,
    author={Bernard, Olivier and Mairet, Francis and Chachuat, Beno{\^i}t},
    title={Modelling of Microalgae Culture Systems with Applications to Control and Optimization},
    bookTitle={Microalgae Biotechnology},
    year={2016},
    publisher={Springer International Publishing},
    address={Cham},
    chapter={3},
    pages={59--87},
    isbn={978-3-319-23808-1},
    doi={10.1007/10\_2014\_287},
}

@inproceedings{otalora2021dynamic,
    author={Ot{\'a}lora, Pablo and Guzm{\'a}n, Jos{\'e} Luis and Berenguel, Manuel and Aci{\'e}n, Francisco Gabriel},
    editor={Gon{\c{c}}alves, Jos{\'e} Alexandre and Braz-C{\'e}sar, Manuel and Coelho, Jo{\~a}o Paulo},
    title={Dynamic Model for the {pH} in a Raceway Reactor Using Deep Learning Techniques},
    booktitle={CONTROLO 2020},
    year={2021},
    publisher={Springer International Publishing},
    address={Cham},
    pages={190--199},
    isbn={978-3-030-58653-9},
    doi = {10.1007/978-3-030-58653-9\_18}
}

@inproceedings{otalora2024modeling,
    author = {Otálora, Pablo and Skogestad, Sigurd and Guzmán, José Luis and Berenguel, Manuel},
    title = {Modeling, Control and Online Optimization of Microalgae-based Biomass Production in Raceway Reactors},
    pages = {235-240},
    booktitle = {12th IFAC Symposium on Advanced Control of Chemical Processes},
    year = {2024}
}

@article{fernandez2016dynamic,
    title = {Dynamic model of an industrial raceway reactor for microalgae production},
    journal = {Algal Research},
    volume = {17},
    pages = {67-78},
    year = {2016},
    issn = {2211-9264},
    doi = {10.1016/j.algal.2016.04.021},
    author = {I. Fernández and Francisco G. Acién and José L. Guzmán and M. Berenguel and J L. Mendoza},
}

@article{rodriguezmiranda2021anew,
    author = {Rodríguez-Miranda, Enrique and Acién, Francisco G. and Guzmán, Jose L. and Berenguel, Manuel and Visioli, Antonio},
    title = {A new model to analyze the temperature effect on the microalgae performance at large scale raceway reactors},
    journal = {Biotechnology and Bioengineering},
    volume = {118},
    number = {2},
    pages = {877-889},
    doi = {10.1002/bit.27617},
    year = {2021}
}

@article{nordio2024abaco2,
    title = {{ABACO}-2: a comprehensive model for microalgae-bacteria consortia validated outdoor at pilot-scale},
    journal = {Water Research},
    volume = {248},
    pages = {120837},
    year = {2024},
    issn = {0043-1354},
    doi = {10.1016/j.watres.2023.120837},
    author = {Rebecca Nordio and Enrique Rodríguez-Miranda and Francesca Casagli and Ana Sánchez-Zurano and José Luis Guzmán and Gabriel Acién}
}

@article{solimeno2019bio_algae,
    title={{BIO\_ALGAE} 2: improved model of microalgae and bacteria consortia for wastewater treatment},
    author={Solimeno, Alessandro and G{\'o}mez-Serrano, Cintia and Aci{\'e}n, Francisco Gabriel},
    journal={Environmental Science and Pollution Research},
    volume={26},
    pages={25855--25868},
    year={2019},
    publisher={Springer},
    doi={10.1007/s11356-019-05824-5}
}

@article{fernandez2014first,
    title={First principles model of a tubular photobioreactor for microalgal production},
    author={Fernandez, Ignacio and Acién, F Gabriel and Berenguel, Manuel and Guzmán, José Luis},
    journal={Industrial \& Engineering Chemistry Research},
    volume={53},
    number={27},
    pages={11121--11136},
    year={2014},
    publisher={ACS Publications},
    doi={10.1021/ie501438r}
}

@article{rodriguezmiranda2020diurnal,
    author = {Rodríguez-Miranda, E. and Guzmán, J. L. and Berenguel, M. and Acién, F. G. and Visioli, A.},
    title = "{Diurnal and nocturnal pH control in microalgae raceway reactors by combining classical and event-based control approaches}",
    journal = {Water Science and Technology},
    volume = {82},
    number = {6},
    pages = {1155-1165},
    year = {2020},
    month = {05},
    issn = {0273-1223},
    doi = {10.2166/wst.2020.260}
}

@article{rodriguezmiranda2019daytime,
    author = {Rodríguez-Miranda, Enrique and Beschi, Manuel and Guzmán, José Luis and Berenguel, Manuel and Visioli, Antonio},
    title = {Daytime/Nighttime Event-Based {PI} Control for the {pH} of a Microalgae Raceway Reactor},
    journal = {Processes},
    volume = {7},
    year = {2019},
    number = {5},
    article-number = {247},
    issn = {2227-9717},
    doi = {10.3390/pr7050247}
}

@article{pawlowski2016event,
    title = {Event-based selective control strategy for raceway reactor: A simulation study},
    journal = {IFAC-PapersOnLine},
    volume = {49},
    number = {7},
    pages = {478-483},
    year = {2016},
    note = {11th IFAC Symposium on Dynamics and Control of Process SystemsIncluding Biosystems DYCOPS-CAB 2016},
    issn = {2405-8963},
    doi = {https://doi.org/10.1016/j.ifacol.2016.07.388},
    author = {A. Pawlowski and I. Frenández and J L. Guzmán and M. Berenguel and F G. Acién and S. Dormido}
}

@article{rodriguezmiranda2021indirect,
    author = {Rodríguez-Miranda, E. and Guzmán, J. L. and Acién, F. G. and Berenguel, M. and Visioli, A.},
    title = {Indirect regulation of temperature in raceway reactors by optimal management of culture depth},
    journal = {Biotechnology and Bioengineering},
    volume = {118},
    number = {3},
    pages = {1186-1198},
    doi = {10.1002/bit.27642},
    year = {2021}
}

@article{mairet2015adaptive,
    title = {Adaptive control of light attenuation for optimizing microalgae production},
    journal = {Journal of Process Control},
    volume = {30},
    pages = {117-124},
    year = {2015},
    note = {CAB/DYCOPS 2013},
    issn = {0959-1524},
    doi = {10.1016/j.jprocont.2015.03.010},
    author = {Francis Mairet and Rafael Muñoz-Tamayo and Olivier Bernard}
}

@article{pataro2023alearning,
    title = {A learning-based model predictive strategy for pH control in raceway photobioreactors with freshwater and wastewater cultivation media},
    journal = {Control Engineering Practice},
    volume = {138},
    pages = {105619},
    year = {2023},
    issn = {0967-0661},
    doi = {10.1016/j.conengprac.2023.105619},
    author = {Igor M.L. Pataro and Juan D. Gil and José L. Guzmán and Manuel Berenguel and João M. Lemos}
}

@article{fernandez2016hierarchical,
    title = {Hierarchical control for microalgae biomass production in photobiorreactors},
    journal = {Control Engineering Practice},
    volume = {54},
    pages = {246-255},
    year = {2016},
    issn = {0967-0661},
    doi = {10.1016/j.conengprac.2016.06.007},
    author = {I. Fernández and M. Berenguel and J L. Guzmán and F G. Acién and G A. {de Andrade} and D J. Pagano}
}

@article{skogestad2003simple,
    title={Simple analytic rules for model reduction and {PID} controller tuning},
    author={Skogestad, Sigurd},
    journal={Journal of process control},
    volume={13},
    number={4},
    pages={291--309},
    year={2003},
    publisher={Elsevier},
    doi = {10.1016/S0959-1524(02)00062-8}
}

@book{camacho2007constrained,
    title={Constrained model predictive control},
    author={Camacho, Eduardo F and Bordons, Carlos and Camacho, Eduardo F and Bordons, Carlos},
    year={2007},
    publisher={Springer}
}

@article{ellis2014economic,
    title = {A tutorial review of economic model predictive control methods},
    journal = {Journal of Process Control},
    volume = {24},
    number = {8},
    pages = {1156-1178},
    year = {2014},
    note = {Economic nonlinear model predictive control},
    issn = {0959-1524},
    doi = {10.1016/j.jprocont.2014.03.010},
    author = {Matthew Ellis and Helen Durand and Panagiotis D. Christofides}
}

@INPROCEEDINGS{rawlings2012fundamentals,
    author={Rawlings, James B. and Angeli, David and Bates, Cuyler N.},
    booktitle={2012 IEEE 51st IEEE Conference on Decision and Control (CDC)}, 
    title={Fundamentals of economic model predictive control}, 
    year={2012},
    volume={},
    number={},
    pages={3851-3861},
    doi={10.1109/CDC.2012.6425822}
}

@article{delahozsiegler2012optimization,
    title = {Optimization of microalgal productivity using an adaptive, non-linear model based strategy},
    journal = {Bioresource Technology},
    volume = {104},
    pages = {537-546},
    year = {2012},
    issn = {0960-8524},
    doi = {10.1016/j.biortech.2011.10.029},
    author = {H. {De la Hoz Siegler} and W.C. McCaffrey and R.E. Burrell and A. Ben-Zvi}
}

@article{pfaffinger2016model,
    title = {Model-based optimization of microalgae areal productivity in flat-plate gas-lift photobioreactors},
    journal = {Algal Research},
    volume = {20},
    pages = {153-163},
    year = {2016},
    issn = {2211-9264},
    doi = {10.1016/j.algal.2016.10.002},
    author = {Christina Evi Pfaffinger and Dennis Schöne and Sascha Trunz and Hannes Löwe and Dirk Weuster-Botz}
}

@INPROCEEDINGS{dewasme2017microalgae,
    author={Dewasme, Laurent and Feudjio Letchindjio, Christian G. and Zuniga, Ixbalank Torres and Vande Wouwer, Alain},
    booktitle={2017 25th Mediterranean Conference on Control and Automation (MED)}, 
    title={Micro-algae productivity optimization using extremum-seeking control}, 
    year={2017},
    volume={},
    number={},
    pages={672-677},
    doi={10.1109/MED.2017.7984195}
}

@article{guzman2020modelado,
    title={Modelado y control de la producción de microalgas en fotobiorreactores industriales},
    volume={18},
    DOI={10.4995/riai.2020.13604},
    number={1}, 
    journal={Revista Iberoamericana de Automática e Informática industrial}, 
    author={Guzmán, J. L. and Acién, F. G. and Berenguel, M.}, 
    year={2020}, 
    month={dic.}, 
    pages={1–18} 
}

@article{bernard2022optimal,
    title = {Optimal optical conditions for Microalgal production in photobioreactors},
    journal = {Journal of Process Control},
    volume = {112},
    pages = {69-77},
    year = {2022},
    issn = {0959-1524},
    doi = {10.1016/j.jprocont.2022.03.001},
    author = {Olivier Bernard and Liu-Di Lu}
}

@Article{ifrim2022model,
    AUTHOR = {Ifrim, George Adrian and Titica, Mariana and Horincar, Georgiana and Antache, Alina and Baicu, Laurențiu and Barbu, Marian and Guzmán, José Luis},
    TITLE = {Model Based Optimal Control of the Photosynthetic Growth of Microalgae in a Batch Photobioreactor},
    JOURNAL = {Energies},
    VOLUME = {15},
    YEAR = {2022},
    NUMBER = {18},
    ARTICLE-NUMBER = {6535},
    ISSN = {1996-1073},
    DOI = {10.3390/en15186535}
}

@article{andersson2019casadi,
    title={CasADi: a software framework for nonlinear optimization and optimal control},
    author={Andersson, Joel AE and Gillis, Joris and Horn, Greg and Rawlings, James B and Diehl, Moritz},
    journal={Mathematical Programming Computation},
    volume={11},
    pages={1--36},
    year={2019},
    publisher={Springer},
    doi={10.1007/s12532-018-0139-4}
}

@article{biegler2009large,
    title = {Large-scale nonlinear programming using IPOPT: An integrating framework for enterprise-wide dynamic optimization},
    journal = {Computers \& Chemical Engineering},
    volume = {33},
    number = {3},
    pages = {575-582},
    year = {2009},
    note = {Selected Papers from the 17th European Symposium on Computer Aided Process Engineering held in Bucharest, Romania, May 2007},
    issn = {0098-1354},
    doi = {10.1016/j.compchemeng.2008.08.006},
    author = {L.T. Biegler and V.M. Zavala}
}

@INPROCEEDINGS{guzman2025microalgae,
    author = {Guzman, Jose Luis and Berenguel, Manuel and Rodríguez Miranda, Enrique and Acien Fernandez, F. Gabriel},
    booktitle={American Control Conference 2025 (ACC2025)}, 
    title={Microalgae production at industrial scale: modelling and control challenges}, 
    year={2025},
    pages={4305-4322},
    doi={10.23919/ACC63710.2025.11107686}
}

@article{casagli2021alba,
    title = {ALBA: A comprehensive growth model to optimize algae-bacteria wastewater treatment in raceway ponds},
    journal = {Water Research},
    volume = {190},
    pages = {116734},
    year = {2021},
    issn = {0043-1354},
    doi = {https://doi.org/10.1016/j.watres.2020.116734},
    author = {Francesca Casagli and Gaetano Zuccaro and Olivier Bernard and Jean-Philippe Steyer and Elena Ficara}
}

@misc{rodriguezmiranda2025acomprehensive,
    author       = {Rodríguez Miranda, Enrique and
                  Guzman, Jose Luis and
                  Acien Fernandez, F. Gabriel and
                  Berenguel, Manuel},
    title        = {A Comprehensive Dynamic Model of Microalgae Production in Open Raceway Systems},
    month        = jun,
    year         = 2025,
    publisher    = {Zenodo},
    doi          = {10.5281/zenodo.15579693},
    url          = {https://doi.org/10.5281/zenodo.15579693},
}

@book{rossiter2017model,
    title={Model-based predictive control: a practical approach},
    author={Rossiter, J Anthony},
    year={2017},
    publisher={CRC press}
}

@article{tang2004comparative,
    title = {Comparative studies on the water evaporation rate from a wetted surface and that from a free water surface},
    journal = {Building and Environment},
    volume = {39},
    number = {1},
    pages = {77-86},
    year = {2004},
    issn = {0360-1323},
    doi = {https://doi.org/10.1016/j.buildenv.2003.07.007},
    author = {Runsheng Tang and Y. Etzion}
}

@misc{weatherbit2025,
    author       = {Weatherbit},
    title        = {Weatherbit {API} Documentation},
    year         = {2025},
    howpublished = {\url{https://www.weatherbit.io/}}
}

\end{document}